\documentclass[journal,draftclsnofoot,12pt,onecolumn]{IEEEtran}
\usepackage{graphicx}
\usepackage{amssymb}
\usepackage{color}
\usepackage{subfigure}
\usepackage{caption}
\renewcommand{\theequation}{\arabic{section}.\arabic{equation}}
\newtheorem{theorem}{Theorem}

\newtheorem{lemma}{Lemma}

\newtheorem{remark}{Remark}

\begin{document}

\title{An Improved Feedback Coding Scheme for the Wire-tap Channel}

\author{Bin~Dai
        and~Yuan~Luo
\thanks{B. Dai is with the
School of Information Science and Technology,
Southwest JiaoTong University, Chengdu 610031, China, and with
the State Key Laboratory of Integrated Services Networks, Xidian University, Xi$'$an, Shaanxi 710071, China,
e-mail: daibin@home.swjtu.edu.cn.}
\thanks{Y. Luo is with the computer science and engineering department, Shanghai Jiao Tong University, Shanghai 200240, China,
Email: luoyuan@cs.sjtu.edu.cn.}
}

\maketitle

\begin{abstract}

The model of wiretap channel (WTC) is important as it constitutes the essence of physical layer security (PLS).
Wiretap channel with noiseless feedback (WTC-NF) is especially interesting as it shows what can be
done when a private feedback is available. The already existing secret key based feedback coding scheme focuses on generating key from the feedback and using this key
to protect part of the transmitted message. It has been shown that this secret key based feedback coding scheme
is only optimal for the degraded WTC-NF, and finding an optimal feedback scheme for the general WTC-NF motivates us to exploit
other uses of the feedback. In this paper, a new feedback coding scheme for the general WTC-NF is proposed, where
the feedback is not only used to generate key, but also used to generate help information which helps the
legitimate parties to improve the communication between them. We show that
the proposed new feedback scheme performs better than the already existing one, and
a binary example is given to further explain the results of this paper.

\end{abstract}

\begin{IEEEkeywords}
Feedback, secrecy capacity, wiretap channel, Wyner-Ziv coding.
\end{IEEEkeywords}

\section{Introduction \label{secI}}

Introducing both reliability and security constraints into a physically degraded broadcast channel,
the transmission over noisy communication channels
was first studied by Wyner \cite{Wy}.
The model investigated in \cite{Wy} is known as the physically degraded WTC, and its
secrecy capacity equals
\begin{eqnarray}\label{c1}
&&C^{d}_{s}=\max_{P(x)}[I(Y_{1};X)-I(Y_{2};X)],
\end{eqnarray}
where $X$, $Y_{1}$ and $Y_{2}$ are the random variables (RVs) representing the channel input, the legal receiver's received \textcolor[rgb]{1.00,0.00,0.00}{signal}, and
the wiretapper's received \textcolor[rgb]{1.00,0.00,0.00}{signal}, respectively. Moreover, in \cite{Wy}, ``physically degraded'' indicates
the existence of a Markov condition $X\rightarrow Y_{1}\rightarrow Y_{2}$.

Wyner's physically degraded model \cite{Wy} was further studied by Csisz$\acute{a}$r and K\"{o}rner \cite{CK},
where the WTC without the degradedness assumption $X\rightarrow Y_{1}\rightarrow Y_{2}$ was investigated.
In \cite{CK}, the authors indicated
that the secrecy capacity of this general WTC equaled
\begin{eqnarray}\label{e1.1}
&&C_{s}=\max_{P(x|u),P(u)}[I(Y_{1};U)-I(Y_{2};U)]^{+},
\end{eqnarray}
where the function $[x]^{+}=\max\{x,0\}$, $U$ is the message random variable (RV),
and $U\rightarrow X\rightarrow (Y_{1},Y_{2})$.
Similar to the classical channel coding theorem proposed by Shannon, the channels studied in \cite{Wy} and \cite{CK} are assumed to be
discrete memoryless. For the WTC with Gaussian channel noise, the secrecy capacities of degraded
and general cases are determined in \cite{hell} and \cite{liang}, respectively.

Later, the WTC has been re-considered by assuming the legal receiver's received signal $Y_{1}$ is accessible to the transmitter,
and this model is known as the wiretap channel with noiseless feedback (WTC-NF). \textcolor[rgb]{1.00,0.00,0.00}{In \cite{AC}, Ahlswede and Cai pointed out that
to enhance the secrecy capacity of the WTC, the best use of the legal receiver's feedback channel output is to generate random bits from it
and use these bits as a key by the transmitter protecting part of the transmitted message. Using this feedback coding scheme, Ahlswede and Cai \cite{AC} provided a
lower bound $R_{s}$ on the secrecy capacity $C^{f}_{s}$
of the general WTC-NF, and it is given by
\begin{eqnarray}\label{e1.2}
&&R_{s}=\max_{P(x|u),P(u)}\min\{[I(Y_{1};U)-I(Y_{2};U)]^{+}+H(Y_{1}|Y_{2},U), I(Y_{1};U)\},
\end{eqnarray}
where $U$ is the message RV defined the same as that in (\ref{e1.1}).
Comparing (\ref{e1.2}) with (\ref{e1.1}), it is obvious that the feedback coding scheme in \cite{AC} enlarges the secrecy capacity of
the WTC. Here note that the secrecy capacity $C^{f}_{s}$ of the general WTC-NF
is still unknown.} However, for the physically degraded case ($X\rightarrow Y_{1}\rightarrow Y_{2}$), Ahlswede and Cai \cite{AC} determined the secrecy capacity,
and it equals
\begin{eqnarray}\label{e1.3}
&&C^{df}_{s}=\max_{P(x)}\min\{I(Y_{1};X)-I(Y_{2};X)+H(Y_{1}|Y_{2},X),I(Y_{1};X)\}.
\end{eqnarray}
The comparison of (\ref{e1.3}) and (\ref{c1}) also indicates that the feedback strategy in \cite{AC}
enlarges the secrecy capacity of the physically degraded WTC.

\textcolor[rgb]{1.00,0.00,0.00}{Based on the work of \cite{AC}, Ardestanizadeh et al. \cite{AFJK}
investigated the wiretap channel with rate limited feedback, where the legal receiver is free to use the noiseless feedback channel to send anything
as he wishes (up to a rate $R_{f}$). For the degraded case ($X\rightarrow Y\rightarrow Z$), they showed that the best choice of the legal receiver
is sending a key through the feedback channel, and if the legal receiver's channel output $Y_{1}$ is sent, the best use of it is to extract a key.
Later, Cohen er al. \cite{cohen} generalized Ardestanizadeh et al.'s work \cite{AFJK} by
considering the WTC with noiseless feedback, and with
causal channel state information (CSI) at both the transmitter and the legitimate receiver. Cohen er al. \cite{cohen} showed that
the transmitted message can be protected by two keys, where one is generated from the noiseless feedback, and the other is generated by the causal CSI.
They further showed that these two keys help to enhance the achievable secrecy rate of the WTC with rate limited feedback investigated in \cite{AFJK}.
Besides the work of \cite{cohen},
other related works in the WTC-NF with CSI are investigated in \cite{dainew2}-\cite{dainew3}.
Here note that for the WTC-NF, the current literature (\cite{AC}-\cite{dainew3}) shows that the secrecy capacity is achieved only
for the degraded case ($Y_{2}$ is a degraded version of $Y_{1}$), i.e.,  Ahlswede and Cai's secret key based feedback coding scheme \cite{AC}
is only optimal for the degraded WTC-NF. Finding the optimal feedback coding scheme for the general WTC-NF needs us to exploit
other uses of the feedback.}

\textcolor[rgb]{1.00,0.00,0.00}{In this paper,
we show that for the WTC-NF, the feedback can not only be used to generate key protecting the transmitted message, but also be used to
generate help information which helps the legitimate parties to improve the communication between them. }
Using this feedback coding scheme, we obtain
a new lower bound $R^{*}_{s}$ on $C^{f}_{s}$, and it is given by
\begin{eqnarray}\label{e1.4}
&&R^{*}_{s}=\max_{P(v|y_{1},u),P(x|u),P(u)}\min\{[I(Y_{1},V;U)-I(Y_{2};U)]^{+}+H(Y_{1}|Y_{2},U),I(Y_{1};U)\},
\end{eqnarray}
where $U$ is the message RV defined the same as those in (\ref{e1.1}) and (\ref{e1.2}),
and \textcolor[rgb]{1.00,0.00,0.00}{the auxiliary RV $V$ denotes an estimation of $U$ and $Y_{1}$}.
Comparing (\ref{e1.4}) with (\ref{e1.2}), we can conclude that Ahlswede and Cai's
lower bound $R_{s}$ is no larger than our new lower bound $R^{*}_{s}$
since
\begin{eqnarray}\label{e1.4xxxe}
&&I(Y_{1};U)\leq I(Y_{1},V;U).
\end{eqnarray}
Moreover, it is easy to see that our new lower bound $R^{*}_{s}$ generalizes Ahlswede and Cai's lower bound $R_{s}$ by considering an additional random variable $V$.
Letting $V$ be a constant, $R^{*}_{s}$ reduces to $R_{s}$.

\textcolor[rgb]{1.00,0.00,0.00}{Here note that the new feedback scheme achieving $R^{*}_{s}$ combines
the Wyner-Ziv coding scheme \cite{ziv} used in the source coding with side information (SC-SI) problem
with the Ahlswede and Cai's coding scheme \cite{AC} for the WTC-NF, and the motivation of this combined new scheme is explained below.
First, we notice that the achievability proof of the
Ahlswede and Cai's achievable secrecy rate $R_{s}$ \cite{AC} for the WTC-NF indicates that $R_{s}$ relies heavily on the
number of the transmission codewords. Next,
we apply the Wyner-Ziv coding scheme to the model of the discrete memoryless channel (DMC) with feedback,
and propose a new feedback coding scheme for the DMC with feedback (formally introduced in Section \ref{secII-2}), which
combines the Wyner-Ziv coding scheme \cite{ziv} used in the SC-SI problem with the block Markov coding
scheme for the feedback systems. We show that this new feedback coding scheme also achieves the capacity of the DMC with feedback, and
find that though this new scheme can not
increase the capacity of the DMC, the total number of the transmission codewords increases. Then it is natural to ask whether the
Wyner-Ziv coding scheme helps to increase the total number of the transmission codewords in the Ahlswede and Cai's coding scheme \cite{AC} for the WTC-NF,
and whether the Wyner-Ziv coding scheme helps to enhance the already existing achievable secrecy rate $R_{s}$ for the WTC-NF. To answer these two questions
motivates us to propose the new feedback scheme of this paper.}

Now the rest of this paper is arranged as below. \textcolor[rgb]{1.00,0.00,0.00}{The necessary mathematical background,
the Wyner-Ziv coding scheme and its application to the DMC with feedback are provided in Section \ref{secII-2}.
The model formulation and the main result (a new achievable secrecy rate for the WTC-NF) are given in Section \ref{secII}.
The proof of the main result is given in Section \ref{secf}.
A special case of the WTC-NF, which is called the non-degraded WTC-NF,
is investigated in Section \ref{secIII}. Numerical results on the binary non-degraded WTC-NF are shown
in Section \ref{secIV}, and a summary of this work is given in Section \ref{secV}.}

\begin{figure}[htb]
\centering
\includegraphics[scale=0.6]{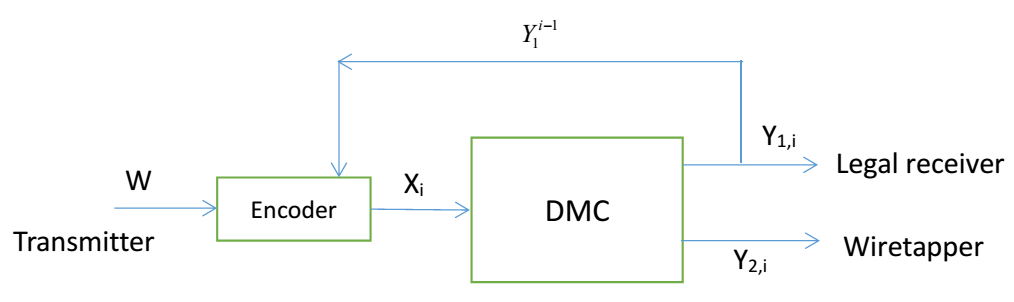}
\caption{The general WTC-NF}
\label{f1}
\end{figure}

\section{\textcolor[rgb]{1.00,0.00,0.00}{Preliminaries}}\label{secII-2}
\setcounter{equation}{0}

\subsection{Notations and Basic Lemmas}

\emph{Notations}: In the remainder of this paper, we use uppercase letters throughout to represent random variables (RVs), lowercase letters for scalars
or observed values of RVs, and calligraphic letters for alphabets. \textcolor[rgb]{1.00,0.00,0.00}{A similar convention is applied to the random vectors and their sample values.
For example, $Y_{1}$ denotes a RV, and $y_{1}$ is a specific value in $\mathcal{Y}_{1}$. Similarly,
$Y_{1}^{N}$ denotes a random $N$-vector $(Y_{1,1},...,Y_{1,N})$,
and $y_{1}^{N}=(y_{1,1},...,y_{1,N})$ is a specific vector value in $\mathcal{Y}_{1}^{N}$
that is the $N$-th Cartesian power of $\mathcal{Y}_{1}$.}
Moreover, we use $P(x)$ as an abbreviation of the event probability $Pr\{X=x\}$,
and notice that the $\log$ function takes base $2$.

An independent identically distributed (i.i.d.)
produced vector $x^{N}$ with respect to (w.r.t.) the probability $P(x)$ is $\epsilon$-typical if for all $x\in \mathcal{X}$,
$$|\frac{\pi_{x^{N}}(x)}{N}-P(x)|\leq \epsilon,$$
where $\pi_{x^{N}}(x)$ denotes the number of $x$ showing up in $x^{N}$.
The strong typical set $T^{N}_{\epsilon}(P(x))$ consists of all typical vectors $x^{N}$.
The following lemmas with respect to $T^{N}_{\epsilon}(P(x))$ will be used in the remainder of this paper.

\begin{lemma}\label{L1}
\textbf{(Covering Lemma \cite{network})}: Let $U^{N}$ and $V^{N}(i)$ ($i\in \mathcal{I}$ and $|\mathcal{I}|\geq 2^{NR}$) be i.i.d.
produced random sequences respectively related to the probabilities $P(u)$ and $P(v)$. The random vector $U^{N}$ is independent of $V^{N}(i)$.
Then there exists $\eta>0$ such that
\begin{eqnarray*}
\lim_{N\rightarrow\infty}P(\forall i\in \mathcal{I},\,\,(U^{N},V^{N}(i))\notin T^{N}_{\eta}(P(u,v)))=0
\end{eqnarray*}
if $R>I(U;V)+\varphi(\eta)$, where $\varphi(\eta)\rightarrow 0$ as $\eta\rightarrow 0$.
\end{lemma}

\begin{lemma}\label{L2}
\textbf{(Packing Lemma \cite{network})}: Let $U^{N}$ and $V^{N}(i)$ ($i\in \mathcal{I}$ and $|\mathcal{I}|\leq 2^{NR}$) be i.i.d.
produced random sequences respectively with respect to the probabilities $P(u)$ and $P(v)$. The random vector $V^{N}(i)$ is independent of $U^{N}$.
Then there exists $\eta>0$ such that
\begin{eqnarray*}
\lim_{N\rightarrow\infty}P(\exists i\in \mathcal{I} \,\,s.t.\,\,(U^{N},V^{N}(i))\in T^{N}_{\eta}(P(u,v)))=0
\end{eqnarray*}
if $R<I(U;V)-\varphi(\eta)$, where $\varphi(\eta)\rightarrow 0$ as $\eta\rightarrow 0$.
\end{lemma}

\begin{lemma}\label{L4}
\textbf{(Balanced coloring lemma \cite[p. 260]{AC})}: For any $\epsilon, \delta>0$ and sufficiently large $N$,
let $U^{N}$, $Y_{1}^{N}$ and $Y_{2}^{N}$ be i.i.d. produced random sequences
respectively with respect to the probabilities $P(u)$, $P(y_{1})$ and $P(y_{2})$. Given $y_{2}^{N}$ and $u^{N}$, let
$T_{P(y_{1}|y_{2},u)}^{N}(y_{2}^{N},u^{N})$ be the conditional strong typical set composed of all $y_{1}^{N}$ satisfying the fact that
$(y_{1}^{N},y_{2}^{N},u^{N})$ are jointly typical. In addition, for $\gamma<|T_{P(y_{1}|y_{2},u)}^{N}(y_{2}^{N},u^{N})|$,
let $\phi$ be a $\gamma$-coloring
$$\phi: T^{N}_{\epsilon}(P(y_{1}))\rightarrow \{1,2,..,\gamma\},$$
and $\phi^{-1}(k)$ ($k\in \{1,2,..,\gamma\}$) be a set composed of all $y_{1}^{N}$ such that
$\phi(y_{1}^{N})=k$ and $y_{1}^{N}\in T_{P(y_{1}|y_{2},u)}^{N}(y_{2}^{N},u^{N})$.
Then we have
\begin{equation}\label{appen1}
|\phi^{-1}(k)|\leq \frac{|T_{P(y_{1}|y_{2},u)}^{N}(y_{2}^{N},u^{N})|(1+\delta)}{\gamma},
\end{equation}
where $k\in \{1,2,..,\gamma\}$.
\end{lemma}
Applying Lemma \ref{L4}, we can conclude that there are at least
\begin{equation}\label{appen2}
\frac{|T_{P(y_{1}|y_{2},u)}^{N}(y_{2}^{N},u^{N})|}{\frac{|T_{P(y_{1}|y_{2},u)}^{N}(y_{2}^{N},u^{N})|(1+\delta)}{\gamma}}=\frac{\gamma}{1+\delta}
\end{equation}
colors and at most $\gamma$ colors
mapped by the conditional strong typical set $T_{P(y_{1}|y_{2},u)}^{N}(y_{2}^{N},u^{N})$.
Letting $\gamma=|T_{P(y_{1}|y_{2},u)}^{N}(y_{2}^{N},u^{N})|$ and using the properties of the conditional
strong typical set \cite{network}, we have
\begin{eqnarray}\label{zhenni1}
&&\gamma=|T_{P(y_{1}|y_{2},u)}^{N}(y_{2}^{N},u^{N})|\geq (1-\epsilon_{1})2^{N(1-\epsilon_{2})H(Y_{1}|U,Y_{2})},
\end{eqnarray}
where $\epsilon_{1}$ and $\epsilon_{2}$ tend to zero while $N\rightarrow \infty$.

\subsection{\textcolor[rgb]{1.00,0.00,0.00}{Wyner-Ziv Coding Scheme for the Lossless SC-SI}}

\begin{figure}[htb]
\centering
\includegraphics[scale=0.5]{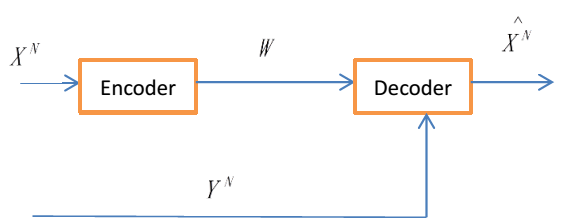}
\caption{\textcolor[rgb]{1.00,0.00,0.00}{The lossless source coding with side information}}
\label{honghai}
\end{figure}

\textcolor[rgb]{1.00,0.00,0.00}{In this subsection, we review the Wyner-Ziv coding scheme.
For the lossless SC-SI shown in Figure \ref{honghai}, the source $X^{N}$ is correlated with a side information $Y^{N}$, and they are
i.i.d. generated according to the probability $P(x,y)$. Using an encoding function
$\phi: \mathcal{X}^{N}\rightarrow \{1,2,...,2^{NR}\}$,
the transmitter compresses $X^{N}$ into an index $W$ taking values in $\{1,2,...,2^{NR}\}$,
and this index $W$ together with the side information $Y^{N}$ are available at the receiver.
The receiver generates a reconstruction sequence $\hat{U}^{N}=\varphi(W,Y^{N})$ by applying a reconstruction function
$\varphi: \{1,2,...,2^{NR}\}\times \mathcal{Y}^{N}\rightarrow \mathcal{U}^{N}$ to the index $W$ and the
side information $Y^{N}$. The goal of the communication is that the reconstruction sequence $\hat{U}^{N}$ is jointly typical with the source
$X^{N}$ according to the probability $P(u|x)\times P(x)$.}

\textcolor[rgb]{1.00,0.00,0.00}{A rate $R$ is said to be achievable if for any $\epsilon>0$, there exists
a sequence of encoding and reconstruction functions $(\phi, \varphi)$ such that}
\begin{eqnarray}\label{kong1-sg1}
&&\textcolor[rgb]{1.00,0.00,0.00}{Pr\{(X^{N},\hat{U}^{N})\notin T^{N}_{\epsilon}(P(x,u))\}\rightarrow 0}
\end{eqnarray}
\textcolor[rgb]{1.00,0.00,0.00}{as $N\rightarrow \infty$.
The following Wyner-Ziv Theorem given in \cite{ziv} characterizes the minimum achievable rate $R$ of this lossless SC-SI problem.}

\begin{theorem}\label{T2-xx}
\textcolor[rgb]{1.00,0.00,0.00}{(Wyner-Ziv Theorem): For the SC-SI, the achievable rate $R$ satisfies}
\begin{eqnarray}\label{trans1x}
&&\textcolor[rgb]{1.00,0.00,0.00}{R\geq \min_{P(u|x)}(I(X;U)-I(Y;U))=\min_{P(u|x)} I(X;U|Y),}
\end{eqnarray}
\textcolor[rgb]{1.00,0.00,0.00}{where $U\rightarrow X\rightarrow Y$ and the alphabet of $U$ is bounded by $|\mathcal{U}|\leq |\mathcal{X}|+1$.}
\end{theorem}

\emph{\textcolor[rgb]{1.00,0.00,0.00}{Achievable coding scheme for Theorem \ref{T2-xx}}}:

\begin{itemize}

\item \textbf{\textcolor[rgb]{1.00,0.00,0.00}{Code-book generation}}:
\textcolor[rgb]{1.00,0.00,0.00}{Generate $2^{N(R+R^{*})}$ i.i.d. sequences $u^{N}$ according to the probability $P(u)$, and index them as
$u^{N}(l)$, where $1\leq l\leq 2^{N(R+R^{*})}$. Partition all of the sequences $u^{N}$ into $2^{NR}$ bins, and each bin contains $2^{NR^{*}}$ sequences.}

\item \textbf{\textcolor[rgb]{1.00,0.00,0.00}{Encoding}}:
\textcolor[rgb]{1.00,0.00,0.00}{Given a source $x^{N}$, the encoder finds an index $l^{*}$ such that $(x^{N},u^{N}(l^{*}))$ are jointly typical. If there is no such index, declare an encoding error.
If there is more than one such index, randomly choose one. Based on the covering lemma (see Lemma \ref{L1}), the encoding error
approaches to zero when}
\begin{eqnarray}\label{kong1}
&&\textcolor[rgb]{1.00,0.00,0.00}{R+R^{*}\geq I(X;U)}.
\end{eqnarray}
\textcolor[rgb]{1.00,0.00,0.00}{Once the sequence $u^{N}(l^{*})$ is chosen,
the encoder sends the bin index $w\in \{1,2,...,2^{NR}\}$ that $u^{N}(l^{*})$ belongs to.}

\item \textbf{\textcolor[rgb]{1.00,0.00,0.00}{Decoding}}:
\textcolor[rgb]{1.00,0.00,0.00}{Upon receiving the bin index $w$, the decoder finds the unique index $\hat{l}^{*}$ belonging to bin $w$
such that $(y^{N},u^{N}(\hat{l}^{*}))$ are jointly typical. If there is no or more than one such index, declare an decoding error.
Based on the packing lemma (see Lemma \ref{L2}), the decoding error
approaches to zero when}
\begin{eqnarray}\label{kong2}
&&\textcolor[rgb]{1.00,0.00,0.00}{R^{*}\leq I(Y;U).}
\end{eqnarray}

\item \textcolor[rgb]{1.00,0.00,0.00}{Combining (\ref{kong1}) with (\ref{kong2}), and minimizing the bound $R\geq I(X;U)-I(Y;U)$, Theorem \ref{T2-xx} is proved.}

\end{itemize}

\subsection{\textcolor[rgb]{1.00,0.00,0.00}{Application of Wyner-Ziv Coding Scheme to the Discrete Memoryless Channel with Feedback}}

\textcolor[rgb]{1.00,0.00,0.00}{It is well known that the feedback does not increase the capacity of a DMC, and hence the classical channel coding method
is enough to achieve the capacity of the DMC with feedback.
In this subsection, we apply the Wyner-Ziv coding scheme introduced in the last subsection to the DMC with noiseless feedback, and propose
a new feedback scheme for the DMC.
We show that this new feedback scheme also achieves the capacity of the DMC with feedback, and compare this new scheme with the
classical channel coding method.}

\begin{figure}[htb]
\centering
\includegraphics[scale=0.5]{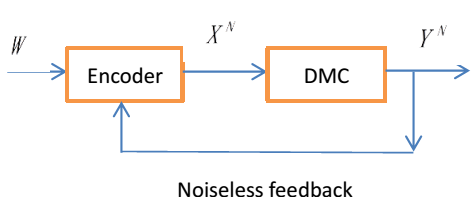}
\caption{\textcolor[rgb]{1.00,0.00,0.00}{The DMC with noiseless feedback}}
\label{huanxi}
\end{figure}

\emph{\textcolor[rgb]{1.00,0.00,0.00}{New coding scheme for the channel with feedback}}:

\textcolor[rgb]{1.00,0.00,0.00}{For the feedback channels, the block Markov coding scheme has been shown to be a useful tool to
improve the receiver's decoding performance. Now we propose a new coding scheme for the DMC with feedback combining
the Wyner-Ziv coding scheme
with the block Markov coding scheme for the feedback systems.}

\textcolor[rgb]{1.00,0.00,0.00}{In the block Markov coding scheme,
the messages are transmitted through $n$ blocks, and the codeword length in each block is $N$.
Define the overall message $w$ transmitted through $n$ blocks as
$w=(w_{1},...,w_{n})$, where $w_{i}$ ($1\leq i\leq n$) is the message for the $i$-th block, and $w_{i}\in\{1,2,...,2^{NR}\}$.
Similar to the construction of $u^{N}$ in Wyner-Ziv coding scheme,
in block $i$ ($1\leq i\leq n$), let $w^{*}_{i}$ and $w^{**}_{i}$ be two indexes respectively
taking values in $\{1,2,...,2^{NR^{*}}\}$ and $\{1,2,...,2^{NR^{**}}\}$.
In addition, for convenience, denote the random vectors $X^{N}$, $Y^{N}$ and $V^{N}$ of block $i$ by
$\bar{X}_{i}$, $\bar{Y}_{i}$ and $\bar{V}_{i}$, respectively.
The vector value is written in lower case letter.}

\begin{itemize}

\item \textbf{\textcolor[rgb]{1.00,0.00,0.00}{Code-book generation}}:
\textcolor[rgb]{1.00,0.00,0.00}{In block $i$, generate $2^{N(R+R^{*})}$ i.i.d. sequences $\bar{x}_{i}$ according to the probability $P(x)$, and index them as
$\bar{x}_{i}(w_{i},w^{*}_{i})$, where $1\leq w_{i}\leq 2^{NR}$ and $1\leq w^{*}_{i}\leq 2^{NR^{*}}$.
Generate $2^{N(R^{*}+R^{**})}$ i.i.d. sequences $\bar{v}_{i}$ according to the probability $P(v|x,y)$, and index them as
$\bar{v}_{i}(w^{*}_{i},w^{**}_{i})$, where $1\leq w^{**}_{i}\leq 2^{NR^{**}}$.
Here note that $\bar{v}_{i}(w^{*}_{i},w^{**}_{i})$ is similar to the codeword $u^{N}$ described in the above Wyner-Ziv coding scheme, and $w^{*}_{i}$
is somewhat like the bin index $w$ of $u^{N}$.}

\item \textbf{\textcolor[rgb]{1.00,0.00,0.00}{Encoding}}:
\textcolor[rgb]{1.00,0.00,0.00}{In block $1$, the transmitter chooses $\bar{x}_{1}(w_{1},1)$ to transmit.
In block $i$ ($2\leq i\leq n$), after receiving the feedback $\bar{y}_{i-1}$, the transmitter tries to choose a $\bar{v}_{i-1}$
such that $(\bar{v}_{i-1},\bar{x}_{i-1},\bar{y}_{i-1})$ are jointly typical. If there is no such $\bar{v}_{i-1}$, declare an encoding error.
If there is more than one such $\bar{v}_{i-1}$, randomly choose one. Based on the covering lemma (see Lemma \ref{L1}), the encoding error
approaches to zero when}
\begin{eqnarray}\label{kong1-dsx}
&&\textcolor[rgb]{1.00,0.00,0.00}{R^{*}+R^{**}\geq I(V;X,Y)}.
\end{eqnarray}
\textcolor[rgb]{1.00,0.00,0.00}{Once the transmitter selects such a $\bar{v}_{i-1}(w^{*}_{i-1},w^{**}_{i-1})$, he chooses $\bar{x}_{i}(w_{i},w^{*}_{i-1})$ for transmission.
In block $n$, after receiving the feedback $\bar{y}_{n-1}$, the transmitter tries to choose a $\bar{v}_{n-1}$
such that $(\bar{v}_{n-1}(w^{*}_{n-1},w^{**}_{n-1}),\bar{x}_{n-1},\bar{y}_{n-1})$ are jointly typical. When successfully decoding such
$\bar{v}_{n-1}(w^{*}_{n-1},w^{**}_{n-1})$, the transmitter picks out $\bar{x}_{n}(1,w^{*}_{n-1})$ for transmission.
The following Figure \ref{huan1} illustrates this encoding procedure.}

\item \textbf{\textcolor[rgb]{1.00,0.00,0.00}{Decoding}}:
\textcolor[rgb]{1.00,0.00,0.00}{The decoding procedure starts from the last block. At block $n$, the receiver chooses a $\bar{x}_{n}$
which is jointly typical with $\bar{y}_{n}$. For the case that more than one or no such $\bar{x}_{n}$ exists, proclaim
an decoding error.
Based on the packing lemma (see Lemma \ref{L2}), this decoding error approaches to zero when}
\begin{eqnarray}\label{app-xxs1}
&&\textcolor[rgb]{1.00,0.00,0.00}{R^{*}\leq I(Y;X).}
\end{eqnarray}
\textcolor[rgb]{1.00,0.00,0.00}{After decoding $\bar{x}_{n}$,
the receiver picks out $w^{*}_{n-1}$ from it. Then
he attempts to select only one $\bar{v}_{n-1}$ such that given $w^{*}_{n-1}$,
$\bar{v}_{n-1}$ is jointly typical with $\bar{y}_{n-1}$ (this is similar to that in the Wyner-Ziv coding scheme \cite{ziv},
where $\bar{y}_{n-1}$ serves as the side information).
For the case that more than one or no such $\bar{v}_{n-1}$ exists,
proclaim an decoding error.
On the basis of Lemma \ref{L2}, this decoding error approaches to zero when}
\begin{eqnarray}\label{app-xxs2}
&&\textcolor[rgb]{1.00,0.00,0.00}{R^{**}\leq I(Y;V).}
\end{eqnarray}
\textcolor[rgb]{1.00,0.00,0.00}{After obtaining such unique $\bar{v}_{n-1}$, the receiver tries to pick
only one $\bar{x}_{n-1}$ which is jointly typical with $(\bar{y}_{n-1},\bar{v}_{n-1})$.
According to Lemma \ref{L2}, this decoding error approaches to zero when}
\begin{eqnarray}\label{app-xxs3}
&&\textcolor[rgb]{1.00,0.00,0.00}{R+R^{*}\leq I(Y,V;X).}
\end{eqnarray}
\textcolor[rgb]{1.00,0.00,0.00}{After decoding $\bar{x}_{n-1}$, the receiver picks out $w_{n-1}$ and $w^{*}_{n-2}$ from it.
Analogously, the receiver decodes the messages $w_{n-2},w_{n-3},...,w_{1}$, and the decoding procedure
is completed. The following Figure \ref{huan2} illustrates the decoding procedure.}

\item \textcolor[rgb]{1.00,0.00,0.00}{Now, substituting (\ref{app-xxs2}) into (\ref{kong1-dsx}), we have}
\begin{eqnarray}\label{app-xxs4}
&&\textcolor[rgb]{1.00,0.00,0.00}{R^{*}\geq I(V;X|Y).}
\end{eqnarray}
\textcolor[rgb]{1.00,0.00,0.00}{Then substituting (\ref{app-xxs4}) into (\ref{app-xxs3}), we get}
\begin{eqnarray}\label{app-xxs5}
&&\textcolor[rgb]{1.00,0.00,0.00}{R\leq I(Y,V;X)-I(V;X|Y)}\nonumber\\
&&\textcolor[rgb]{1.00,0.00,0.00}{=H(X)-H(X|Y,V)-H(X|Y)+H(X|Y,V)=I(X;Y).}
\end{eqnarray}
\textcolor[rgb]{1.00,0.00,0.00}{From (\ref{app-xxs5}), it is easy to see that our new feedback scheme also achieves the capacity of the DMC with feedback.}

\end{itemize}

\emph{\textcolor[rgb]{1.00,0.00,0.00}{Comparison of this new scheme with the classical channel coding method}}:

\textcolor[rgb]{1.00,0.00,0.00}{From the above description of our new scheme, it is easy to see that the total number of the codeword $x^{N}$ is $2^{N(R+R^{*})}$, which is upper bounded
by $2^{NI(X;V,Y)}$. Since feedback does not increase the capacity of a DMC, Shannon's channel coding method \cite{network} is enough to achieve the capacity of the DMC with feedback. In the classical channel coding method, there exists a one-to-one mapping between the transmitted message $w$ and the codeword $x^{N}$,
hence the total number of $X^{N}$ is $2^{NR}$, and it is upper bounded by $2^{NI(X;Y)}$. Since $I(X;V,Y)\geq I(X;Y)$,
our new scheme generates more codewords $x^{N}$ than the classical channel coding method does.
In Section \ref{secf}, we will show that the number of $x^{N}$ has great influence on the achievable secrecy rate of the WTC-NF.
Due to the reason that our new scheme generates more $x^{N}$ than the classical channel coding method does, we guess that
combining this new feedback scheme for the DMC with Ahlswede and Cai's secret key based feedback coding scheme \cite{AC} may achieve a larger secrecy rate
for the WTC-NF.}

\begin{figure}[htb]
\centering
\includegraphics[scale=0.5]{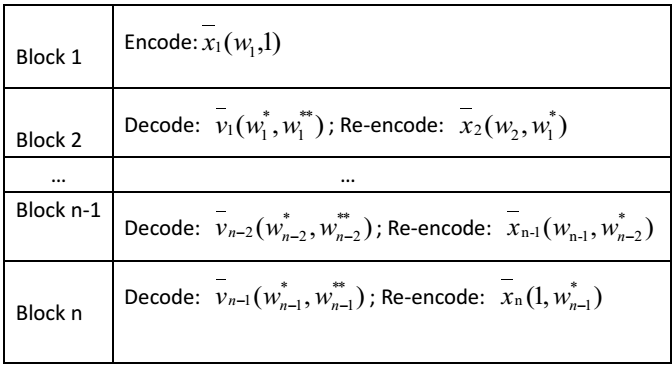}
\caption{\textcolor[rgb]{1.00,0.00,0.00}{Encoding procedure of our new scheme for the DMC with feedback}}
\label{huan1}
\end{figure}

\begin{figure}[htb]
\centering
\includegraphics[scale=0.5]{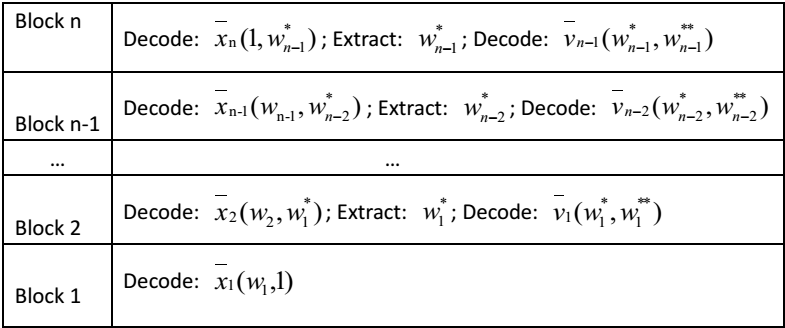}
\caption{\textcolor[rgb]{1.00,0.00,0.00}{Decoding procedure of our new scheme for the DMC with feedback}}
\label{huan2}
\end{figure}

\section{Model Formulation and the Main Result}\label{secII}
\setcounter{equation}{0}

The discrete memoryless WTC-NF consists of one input $x^{N}$, two outputs $y_{1}^{N}$, $y_{2}^{N}$, and satisfies
\begin{equation}\label{shitu1}
P(y_{1}^{N},y_{2}^{N}|x^{N})=\prod_{i=1}^{\textcolor[rgb]{1.00,0.00,0.00}{N}}P(y_{1,i},y_{2,i}|x_{i}),
\end{equation}
where $x_{i}\in \mathcal{X}$, $y_{1,i}\in \mathcal{Y}_{1}$ and $y_{2,i}\in \mathcal{Y}_{2}$.

Let $W$ be the transmission message, its value belongs to
the alphabet $\mathcal{W}=\{1,2,...,M\}$, and
$Pr\{W=i\}=\frac{1}{M}$ for $i\in \mathcal{W}$.
Owing to feedback, the transmitter generates the time-$t$ channel input $X_{t}$ as a function of the message $W$ and
the previously obtained channel outputs $Y_{1,1}$,...,$Y_{1,t-1}$, i.e.,
\begin{equation}\label{e202}
X_{t}=f_{t}(W,Y_{1}^{t-1})
\end{equation}
for some stochastic encoding function $f_{t}$ ($1\leq t\leq N$).

After $N$ channel uses, the legal receiver decodes $W$. Namely, the legal receiver generates the guess
$$\hat{W}=\psi(Y_{1}^{N}),$$
where $\psi$ is the legal receiver's decoding function.
The average probability of the legal receiver's decoding error of the WTC-NF is denoted as
\begin{equation}\label{e204}
P_{e}=\frac{1}{M}\sum_{w\in \mathcal{W}}Pr\{\psi(y_{1}^{N})\neq w|w\,\,\mbox{sent}\}.
\end{equation}
The \textcolor[rgb]{1.00,0.00,0.00}{equivocation rate} of the WTC-NF is defined by
\begin{equation}\label{e205.ff}
\Delta=\frac{1}{N}H(W|Y_{2}^{N}).
\end{equation}

For the WTC-NF, we define an achievable secrecy rate $R$ as follows.
Given a positive number $R$, if for arbitrarily small $\epsilon$,
there exists a pair of channel encoder and decoder
with parameters $M$, $N$ and $P_{e}$ satisfying
\begin{eqnarray}\label{e205}
&&\textcolor[rgb]{1.00,0.00,0.00}{\frac{\log M}{N}\geq R-\epsilon,} \label{e205-1}\\
&&\textcolor[rgb]{1.00,0.00,0.00}{\Delta\geq R-\epsilon,} \label{e205-2}\\
&&\textcolor[rgb]{1.00,0.00,0.00}{P_{e}\leq \epsilon,}\label{e205-3}
\end{eqnarray}
$R$ is called an achievable secrecy rate.
The secrecy capacity $C^{f}_{s}$ is the supremum of achievable secrecy rates for the WTC-NF, and a new lower bound on $C^{f}_{s}$
is shown in Theorem \ref{T2}.

\begin{theorem}\label{T2}
$C^{f}_{s}\geq R^{*}_{s}$,
where
\begin{eqnarray}\label{trans1x}
&&R^{*}_{s}=\max_{P(v|u,y_{1}),P(x|u),P(u)}\min\{[I(Y_{1},V;U)-I(Y_{2};U)]^{+}+H(Y_{1}|Y_{2},U),I(Y_{1};U)\},
\end{eqnarray}
the joint distribution is denoted by
\begin{eqnarray}\label{trans2}
&&P(u,v,x,y_{1},y_{2})=P(v|u,y_{1})P(y_{1},y_{2}|x)P(u,x),
\end{eqnarray}
and the alphabets of $U$ and $V$ are upper bounded by $|\mathcal{U}|\leq |\mathcal{X}|+1$ and $|\mathcal{V}|\leq |\mathcal{X}|+2$, respectively.
\end{theorem}
\begin{IEEEproof}
The bounds on the alphabets of $U$ and $V$ are due to the well known support lemma \cite[pp. 631-633]{network}, and we omit the proof here.
\textcolor[rgb]{1.00,0.00,0.00}{The coding scheme in the proof of Theorem \ref{T2} combines the new coding scheme for the DMC
with feedback shown in Section \ref{secII-2} with
the already existing scheme \cite{AC} for the WTC-NF, and it can be briefly illustrated as follows.
First, in each block, we split the transmitted message into two sub-messages: one is encoded exactly the same
as that in Wyner's wiretap channel \cite{Wy}, and the other is encrypted by a key generated from the feedback.
Then, viewing the two processed sub-messages as a new message $W$, the coding scheme in the proof of Theorem \ref{T2}
is along the lines of the encoding and decoding procedures described in the new coding scheme for the DMC with feedback (see Section \ref{secII-2}).
Here note that
as we have shown in Section \ref{secII-2}, the total rate $R$ does not exceed the main channel capacity, i.e., $R\leq I(Y_{1};U)$.
On the other hand, in each block, since the first sub-message is encoded along the lines of the coding scheme for Wyner's wiretap channel \cite{Wy},
following the equivocation analysis of \cite{Wy} and noticing that the total number of the codewords can be bounded by
(\ref{app-xxs3}) (replacing $X$ by $U$, and $Y$ by $Y_{1}$),
the rate of the first sub-message is upper bounded by $[I(Y_{1},V;U)-I(Y_{2};U)]^{+}$.
For the second sub-message, its rate equals to the rate of the key, and according to the secret key based feedback coding scheme \cite{AC},
the rate of the key is upper bounded by $H(Y_{1}|Y_{2},U)$. Hence the total rate $R$ can be upper bounded by $R\leq [I(Y_{1},V;U)-I(Y_{2};U)]^{+}+H(Y_{1}|Y_{2},U)$.
Combining $R\leq I(Y_{1};U)$ with
$R\leq [I(Y_{1},V;U)-I(Y_{2};U)]^{+}+H(Y_{1}|Y_{2},U)$, and maximizing these two bounds,
$R^{*}_{s}$ is obtained.} The
detail of the proof is in Section \ref{secf}.

\end{IEEEproof}

\begin{remark}\label{R1.1}

Several notes are given below.

\begin{itemize}

\item In \cite{AC}, Ahlswede and Cai also provide an upper bound $C^{f-out}_{s}$ on $C^{f}_{s}$, and it is characterized by
\begin{eqnarray}\label{trans1.xxr}
&&C^{f-out}_{s}=\max_{P(x,u)}\min\{H(Y_{1}|Y_{2}),I(Y_{1};U)\},
\end{eqnarray}
where the joint distribution is denoted by
\begin{eqnarray}\label{trans2.xxr}
&&P(u,x,y_{1},y_{2})=P(y_{1},y_{2}|x)P(u,x).
\end{eqnarray}

\item \textcolor[rgb]{1.00,0.00,0.00}{For the degraded case $X\rightarrow Y_{1}\rightarrow Y_{2}$,
the upper bound $C^{f-out}_{s}$ reduces to
\begin{eqnarray}\label{trans1.xxr-1}
&&C^{f-out}_{s}=\max_{P(x)}\min\{H(Y_{1}|Y_{2}),I(Y_{1};X)\},
\end{eqnarray}
which is consistent with Ahlswede and Cai's secret key based lower bound $R_{s}$ (see (\ref{e1.3})), and this is because
\begin{eqnarray}\label{trans2.xxr-2}
&&I(Y_{1};X)-I(Y_{2};X)+H(Y_{1}|Y_{2},X)=H(X|Y_{2})-H(X|Y_{1})+H(Y_{1}|Y_{2},X)\nonumber\\
&&\stackrel{(1)}=H(X|Y_{2})-H(X|Y_{1},Y_{2})+H(Y_{1}|Y_{2},X)\nonumber\\
&&=I(X;Y_{1}|Y_{2})+H(Y_{1}|Y_{2},X)=H(Y_{1}|Y_{2}),
\end{eqnarray}
where (1) follows from $X\rightarrow Y_{1}\rightarrow Y_{2}$.
Since $R_{s}$ meets the upper bound,
Ahlswede and Cai's secret key based feedback coding scheme is optimal for this degraded WTC-NF, and hence there is no need to further use
$V$ (an estimation of $U$ and $Y_{1}$) to enhance the lower bound $R_{s}$, i.e., for this degraded case, our new lower bound $R^{*}_{s}$ reduces to the secret key based
lower bound $R_{s}$.}

\end{itemize}
\end{remark}

\section{Proof of Theorem \ref{T2}}\label{secf}
\setcounter{equation}{0}

\textcolor[rgb]{1.00,0.00,0.00}{Ahlswede and Cai's secret key based feedback coding scheme \cite{AC}
has been combined with the
new coding scheme for the DMC with feedback provided in Section \ref{secII-2}
to show the achievability of Theorem \ref{T2}.
Now the remainder of this section is organized as follows. The coding strategy is introduced in Subsection \ref{appen1-xx},
and the equivocation analysis is given in Subsection \ref{appen1-xxx}.}

\subsection{Coding Strategy\label{appen1-xx}}

\emph{Definitions and notations}:
\begin{itemize}

\item \textcolor[rgb]{1.00,0.00,0.00}{Similar to the new coding scheme for the DMC with feedback (see Section \ref{secII-2}), assume that
the transmission is through $n$ blocks, and the codeword length in each block is $N$.}

\item The overall message $W$ consists of $n$ components
($W=(W_{1},...,W_{n})$), and
each component $W_{j}$ ($j\in\{1,2,...,n\}$) is the message transmitted in block $j$. The value of $W_{j}$ belongs to
the set $\{1,...,2^{NR}\}$. Then further split $W_{j}$ into two parts $W_{j}=(\textcolor[rgb]{1.00,0.00,0.00}{W_{1,j},W_{2,j}})$, and the values of
$\textcolor[rgb]{1.00,0.00,0.00}{W_{1,j}}$
and $\textcolor[rgb]{1.00,0.00,0.00}{W_{2,j}}$ respectively belong to the sets $\{1,...,2^{NR_{1}}\}$ and $\{1,...,2^{NR_{2}}\}$. Here notice that $R_{1}+R_{2}=R$.

\item Analogously, the randomly produced $W^{'}$, which is used to confuse the wiretapper \footnote{\textcolor[rgb]{1.00,0.00,0.00}{The idea of using random
messages to confuse the wiretapper is exactly the same as the random binning technique used in Wyner's wiretap channel \cite{Wy}, where
this randomly produced message is analogous to the randomly chosen bin index used in the random binning scheme.}}, also consists of $n$ components
($W^{'}=(W^{'}_{1},...,W^{'}_{n})$), and the component $W^{'}_{j}$ ($j\in\{1,2,...,n\}$) is transmitted in block $j$.
\textcolor[rgb]{1.00,0.00,0.00}{Here note that $W^{'}_{j}$ is uniformly drawn from the set $\{1,...,2^{NR^{'}}\}$, i.e.,
$Pr\{W^{'}_{j}=i\}=2^{-NR^{'}}$, where $i\in \{1,...,2^{NR^{'}}\}$.}

\item The help information $W^{*}$, which is used to ameliorate the legal receiver's decoding performance, consists of
$n$ components ($W^{*}=(W^{*}_{1},...,W^{*}_{n})$), and the value of $W^{*}_{j}$ ($j\in\{1,2,...,n\}$)
belongs to the set $\{1,...,2^{NR^{*}}\}$.

\item In block $i$ ($1\leq i\leq n$), the random vectors $X^{N}$, $Y_{1}^{N}$, $Y_{2}^{N}$, $U^{N}$ and $V^{N}$ are denoted by
$\bar{X}_{i}$, $\bar{Y}_{1,i}$, $\bar{Y}_{2,i}$, $\bar{U}_{i}$ and $\bar{V}_{i}$, respectively.
In addition, let $X^{n}=(\bar{X}_{1},...,\bar{X}_{n})$ be a collection of the random vectors $X^{N}$ for all blocks.
Analogously, we have $Y_{1}^{n}=(\bar{Y}_{1,1},...,\bar{Y}_{1,n})$, $Y_{2}^{n}=(\bar{Y}_{2,1},...,\bar{Y}_{2,n})$,
$U^{n}=(\bar{U}_{1},...,\bar{U}_{n})$ and $V^{n}=(\bar{V}_{1},...,\bar{V}_{n})$. The vector value is written in lower case letter.
\end{itemize}

\emph{Code-book generation}:
\begin{itemize}

\item In block $i$ ($1\leq i\leq n$),
randomly produce $2^{N(R_{1}+R_{2}+R^{'}+R^{*})}$ i.i.d. codewords $\bar{u}_{i}$ on the basis of $P(u)$, and label
them as $\bar{u}_{i}(\textcolor[rgb]{1.00,0.00,0.00}{w_{1,i},w_{2,i}},w^{'}_{i},w^{*}_{i-1})$
\footnote{\textcolor[rgb]{1.00,0.00,0.00}{Here we can also say $\bar{u}_{i}$ is a function of $w_{1,i}$, $w_{2,i}$, $w^{'}_{i}$ and $w^{*}_{i-1}$}}, where
$\textcolor[rgb]{1.00,0.00,0.00}{w_{1,i}}\in \{1,2,...,2^{NR_{1}}\}$,
$\textcolor[rgb]{1.00,0.00,0.00}{w_{2,i}}\in \{1,2,...,2^{NR_{2}}\}$, $w^{'}_{i}\in \{1,2,...,2^{NR^{'}}\}$ and $w^{*}_{i-1}\in \{1,2,...,2^{NR^{*}}\}$.

\item For each possible value of $\bar{u}_{i}(\textcolor[rgb]{1.00,0.00,0.00}{w_{1,i},w_{2,i}},w^{'}_{i},w^{*}_{i-1})$ and
$\bar{y}_{1,i}$, randomly produce $2^{N\tilde{R}}$ i.i.d.
codewords $\bar{v}_{i}$
on the basis of $P(v|u,y_{1})$. Then label these $\bar{v}_{i}$ as $\bar{v}_{i}(w^{*}_{i},w^{**}_{i})$, where
$w^{*}_{i}\in \{1,2,...,2^{NR^{*}}\}$ and $w^{**}_{i}\in \{1,2,...,2^{N(\tilde{R}-R^{*})}\}$.

\item For a given $\bar{u}_{i}$, the transmitted sequence $\bar{x}_{i}$ is i.i.d. produced on the basis of the probability $P(x|u)$.

\end{itemize}

\emph{Encoding strategy}:
\begin{itemize}

\item For block $1$, the transmitter chooses $\bar{u}_{1}(w_{1,1},w_{2,1}=1,w^{'}_{1},w^{*}_{0}=1)$ for transmission.

\item For block $i$ ($i\in\{2,3,...,n-1\}$), before choosing $\bar{u}_{i}$, produce a mapping
$g_{i}: \bar{y}_{1,i-1}\rightarrow \{1,2,...,2^{NR_{2}}\}$ (\textcolor[rgb]{1.00,0.00,0.00}{here note that this mapping is generated
exactly the same as that in \cite{AC}}). Based on this mapping, generate a RV
$K_{i}=g_{i}(\bar{Y}_{1,i-1})$ taking values in $\{1,2,...,2^{NR_{2}}\}$, and $Pr\{K_{i}=j\}=2^{-NR_{2}}$ for $j\in \{1,2,...,2^{NR_{2}}\}$.
The RV $K_{i}$ is used as a secret key and it is not known \textcolor[rgb]{1.00,0.00,0.00}{to} the wiretapper,
and $K_{i}$ is independent of the real transmitted messages $W_{1,i}$ and $W_{2,i}$ for block $i$.
Notice that $k_{i}=g_{i}(\bar{y}_{1,i-1})\in \{1,2,...,2^{NR_{2}}\}$ is a real value of $K_{i}$.
The mapping $g_{i}$ is known \textcolor[rgb]{1.00,0.00,0.00}{to} all parties.
Once the transmitter receives $\bar{y}_{1,i-1}$,
he tries to choose a $\bar{v}_{i-1}$ such that
$(\bar{v}_{i-1},\bar{u}_{i-1},\bar{y}_{1,i-1})$ are jointly typical.
For the case that more than one $\bar{v}_{i-1}$ exist,
pick one at random; if no such $\bar{v}_{i-1}$ exists, proclaim an encoding error.
On the basis of Lemma \ref{L1}, the encoding error approaches to zero when
\begin{eqnarray}\label{app1}
&&\tilde{R}\geq I(U,Y_{1};V).
\end{eqnarray}
Once the transmitter selects such a $\bar{v}_{i-1}(w^{*}_{i-1},w^{**}_{i-1})$,
he picks out $\bar{u}_{i}(\textcolor[rgb]{1.00,0.00,0.00}{w_{1,i},w_{2,i}}\oplus k_{i},w^{'}_{i},w^{*}_{i-1})$ for transmission.

\item At block $n$, after receiving the feedback $\bar{y}_{1,n-1}$,
the transmitter tries to select a $\bar{v}_{n-1}(w^{*}_{n-1},w^{**}_{n-1})$ such that
$(\bar{v}_{n-1}(w^{*}_{n-1},w^{**}_{n-1}), \bar{u}_{n-1}, \bar{y}_{1,n-1})$ are jointly typical. When successfully decoding
such $\bar{v}_{n-1}(w^{*}_{n-1},w^{**}_{n-1})$, the transmitter picks out $\bar{u}_{n}(1,1,1,w^{*}_{n-1})$ for transmission.

\end{itemize}

\emph{Decoding strategy}:

The decoding procedure starts from the last block, i.e., block $n$. At block $n$, the legal receiver chooses a $\bar{u}_{n}$
which is jointly typical with $\bar{y}_{1,n}$. For the case that more than one or no such $\bar{u}_{n}$ exists, proclaim
a decoding error.
On the basis of Lemma \ref{L2}, this kind of decoding error approaches to zero when
\begin{eqnarray}\label{app2}
&&R^{*}\leq I(Y_{1};U).
\end{eqnarray}
After decoding $\bar{u}_{n}$,
the legal receiver picks out $w^{*}_{n-1}$ from it. Then
he attempts to select only one $\bar{v}_{n-1}$ such that given $w^{*}_{n-1}$,
$\bar{v}_{n-1}$ is jointly typical with $\bar{y}_{1,n-1}$. For the case that more than one or no such $\bar{v}_{n-1}$ exists,
proclaim a decoding error.
On the basis of Lemma \ref{L2}, this kind of decoding error approaches to zero when
\begin{eqnarray}\label{app3}
&&\tilde{R}-R^{*}\leq I(Y_{1};V).
\end{eqnarray}
After obtaining such unique $\bar{v}_{n-1}$, the legal receiver attempts to pick
only one $\bar{u}_{n-1}$ which is jointly typical with $(\bar{y}_{1,n-1},\bar{v}_{n-1})$.
According to Lemma \ref{L2}, this kind of decoding error approaches to zero when
\begin{eqnarray}\label{app4}
&&R_{1}+R_{2}+R^{'}+R^{*}\leq I(Y_{1},V;U).
\end{eqnarray}
After decoding $\bar{u}_{n-1}$, the legal receiver picks out $\textcolor[rgb]{1.00,0.00,0.00}{w_{1,n-1}}$,
$\textcolor[rgb]{1.00,0.00,0.00}{w_{2,n-1}}\oplus k_{n-1}$, $w^{*}_{n-2}$ from it.
Notice that the legal receiver has full knowledge of $k_{n-1}=g_{n-1}(\bar{y}_{1,n-2})$, and hence he gets the message
$w_{n-1}=(\textcolor[rgb]{1.00,0.00,0.00}{w_{1,n-1},w_{2,n-1}})$. Analogously, the legal receiver decodes the messages
$w_{n-2},w_{n-3},...,w_{1}$, and the decoding procedure
is completed. For convenience, the encoding and decoding strategies are
illustrated by the following Figure \ref{t1} and Figure \ref{t2}, respectively.

\begin{figure}[htb]
\centering
\includegraphics[scale=0.5]{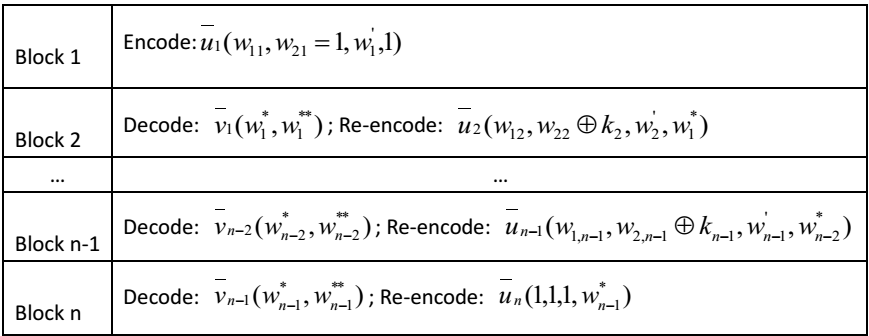}
\caption{\textcolor[rgb]{1.00,0.00,0.00}{The encoding strategy for all blocks}}
\label{t1}
\end{figure}

\begin{figure}[htb]
\centering
\includegraphics[scale=0.5]{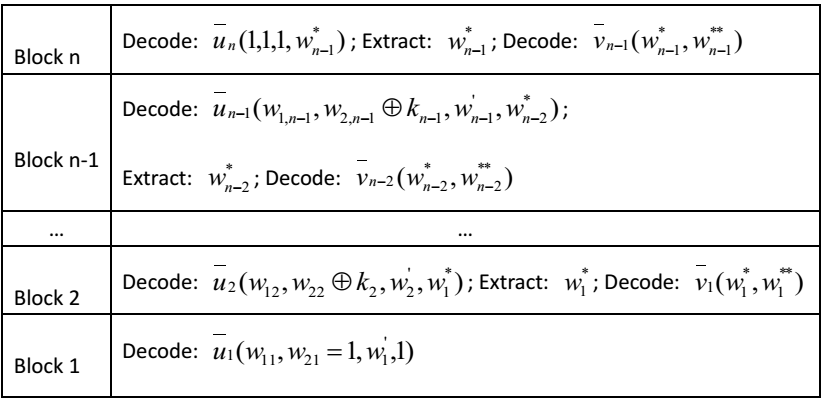}
\caption{\textcolor[rgb]{1.00,0.00,0.00}{The decoding strategy for all blocks}}
\label{t2}
\end{figure}

\subsection{Equivocation Analysis\label{appen1-xxx}}

The overall equivocation $\Delta$, which is denoted by $\Delta=\frac{1}{nN}H(W|Y_{2}^{n})$, equals
\begin{eqnarray}\label{app9}
&&\Delta\stackrel{(a)}=\frac{1}{nN}(H(\tilde{W}_{1}|Y_{2}^{n})+H(\tilde{W}_{2}|Y_{2}^{n},\tilde{W}_{1})),
\end{eqnarray}
where (a) is due to the definitions \textcolor[rgb]{1.00,0.00,0.00}{$\tilde{W}_{1}=(W_{1,1},...,W_{1,n})$ and $\tilde{W}_{2}=(W_{2,1},...,W_{2,n})$}.

The first term $H(\tilde{W}_{1}|Y_{2}^{n})$ of (\ref{app9}) follows that
\begin{eqnarray}\label{app10}
&&H(\tilde{W}_{1}|Y_{2}^{n})=H(\tilde{W}_{1},Y_{2}^{n})-H(Y_{2}^{n})\nonumber\\
&&=H(\tilde{W}_{1},Y_{2}^{n},U^{n})-H(U^{n}|\tilde{W}_{1},Y_{2}^{n})-H(Y_{2}^{n})\nonumber\\
&&\stackrel{(b)}=H(U^{n})-H(U^{n}|\tilde{W}_{1},Y_{2}^{n})-I(U^{n};Y_{2}^{n})\nonumber\\
&&\stackrel{(c)}=(n-1)NR_{1}+(n-2)NR_{2}+(n-1)NR^{'}+(n-1)NR^{*}-nNI(U;Y_{2})-H(U^{n}|\tilde{W}_{1},Y_{2}^{n})\nonumber\\
&&\stackrel{(d)}\geq (n-1)NR_{1}+(n-2)NR_{2}+(n-1)NR^{'}+(n-1)NR^{*}-nNI(U;Y_{2})-nN\epsilon_{3},
\end{eqnarray}
where (b) is implied by $H(\tilde{W}_{1}|U^{n})=0$, (c) is due to the construction of $U^{n}$ and
the channel is memoryless,
and (d) is due to that given $\tilde{w}_{1}$ and $y_{2}^{n}$, the wiretapper attempts to select only one $u^{n}$ that is jointly typical with his own received
\textcolor[rgb]{1.00,0.00,0.00}{signals} $y_{2}^{n}$, and
implied by Lemma \ref{L2}, we can conclude that
the wiretapper's decoding error approaches to zero if
\begin{eqnarray}\label{app11}
&&R_{2}+R^{'}+R^{*}\leq I(Y_{2};U),
\end{eqnarray}
then applying Fano's lemma, $\frac{1}{nN}H(U^{n}|\tilde{W}_{1},Y_{2}^{n})\leq \epsilon_{3}$ is obtained, where $\epsilon_{3}\rightarrow 0$
while $n, N\rightarrow \infty$.

Moreover, the second term $H(\tilde{W}_{2}|Y_{2}^{n},\tilde{W}_{1})$ of (\ref{app9}) follows that
\begin{eqnarray}\label{app12}
&&H(\tilde{W}_{2}|Y_{2}^{n},\tilde{W}_{1})\nonumber\\
&&\geq \sum_{i=2}^{n-1}H(\textcolor[rgb]{1.00,0.00,0.00}{W_{2,i}}|Y_{2}^{n},\tilde{W}_{1},\textcolor[rgb]{1.00,0.00,0.00}{W_{2,1}=1,...,W_{2,i-1},W_{2,i}}\oplus K_{i})\nonumber\\
&&\stackrel{(e)}=\sum_{i=2}^{n-1}H(\textcolor[rgb]{1.00,0.00,0.00}{W_{2,i}}|\bar{Y}_{2,i-1},\textcolor[rgb]{1.00,0.00,0.00}{W_{2,i}}\oplus K_{i})\nonumber\\
&&\geq \sum_{i=2}^{n-1}H(\textcolor[rgb]{1.00,0.00,0.00}{W_{2,i}}|\bar{Y}_{2,i-1},\bar{U}_{i-1},\textcolor[rgb]{1.00,0.00,0.00}{W_{2,i}}\oplus K_{i})\nonumber\\
&&=\sum_{i=2}^{n-1}H(K_{i}|\bar{Y}_{2,i-1},\bar{U}_{i-1},\textcolor[rgb]{1.00,0.00,0.00}{W_{2,i}}\oplus K_{i})\nonumber\\
&&\stackrel{(f)}=\sum_{i=2}^{n-1}H(K_{i}|\bar{Y}_{2,i-1},\bar{U}_{i-1})\nonumber\\
&&\stackrel{(g)}\geq (n-2)(\log\frac{1-\epsilon_{1}}{1+\delta}+N(1-\epsilon_{2})H(Y_{1}|U,Y_{2})),
\end{eqnarray}
where (e) is due to the Markov chain $W_{2,i}\rightarrow (\bar{Y}_{2,i-1},W_{2,i}\oplus K_{i})
\rightarrow (\tilde{W}_{1},\textcolor[rgb]{1.00,0.00,0.00}{W_{2,1},...,W_{2,i-1}},\\ \bar{Y}_{2,1},...,\bar{Y}_{2,i-2},
\bar{Y}_{2,i},...,\bar{Y}_{2,n})$, (f) is implied by
$K_{i}\rightarrow (\bar{Y}_{2,i-1},\bar{U}_{i-1})\rightarrow \textcolor[rgb]{1.00,0.00,0.00}{W_{2,i}}\oplus K_{i}$,
and (g) is implied by Lemma \ref{L4} that given $\bar{y}_{2,i-1}$ and $\bar{u}_{i-1}$, there are at least
$\frac{\gamma}{1+\delta}$ colors (see (\ref{appen2})), which indicates that
\begin{eqnarray}\label{zhenni2}
&&H(K_{i}|\bar{Y}_{2,i-1},\bar{U}_{i-1})
\geq \log\frac{\gamma}{1+\delta},
\end{eqnarray}
then substituting (\ref{zhenni1}) into (\ref{zhenni2}), we get
\begin{eqnarray}\label{zhenni3}
&&H(K_{i}|\bar{Y}_{2,i-1},\bar{U}_{i-1})
\geq \log\frac{1-\epsilon_{1}}{1+\delta}+N(1-\epsilon_{2})H(Y_{1}|U,Y_{2}),
\end{eqnarray}
where $\epsilon_{1}$, $\epsilon_{2}$ and $\delta$ approach to $0$ as $N$ goes to infinity.

Substituting (\ref{app10}) and (\ref{app12}) into (\ref{app9}), we have
\begin{eqnarray}\label{app13}
&&\Delta\geq \frac{n-1}{n}(R_{1}+R^{'}+R^{*})+\frac{n-2}{n}R_{2}-I(Y_{2};U)-\epsilon_{3}\nonumber\\
&&+\frac{n-2}{nN}\log\frac{1-\epsilon_{1}}{1+\delta}+\frac{n-2}{n}(1-\epsilon_{2})H(Y_{1}|Y_{2},U).
\end{eqnarray}
The bound (\ref{app13}) indicates that if
\begin{eqnarray}\label{app14}
&&R^{'}+R^{*}\geq I(Y_{2};U)-H(Y_{1}|Y_{2},U),
\end{eqnarray}
$\Delta\geq R_{1}+R_{2}-\epsilon$ can be proved
by selecting sufficiently large $n$ and $N$.

Now combining (\ref{app1}) with (\ref{app3}), we get
\begin{eqnarray}\label{app14.x1}
&&R^{*}\geq I(U,Y_{1};V)-I(Y_{1};V)=I(V;U|Y_{1}).
\end{eqnarray}
Then implied by (\ref{app14.x1}) and (\ref{app4}), we can conclude that
\begin{eqnarray}\label{zhenni4}
&&R_{1}+R_{2}+R^{'}\leq I(Y_{1},V;U)-I(V;U|Y_{1})=I(U;Y_{1}).
\end{eqnarray}
Next, implied by (\ref{app14}) and (\ref{app4}), we have
\begin{eqnarray}\label{zhenni5}
&&R_{1}+R_{2}\leq I(Y_{1},V;U)-I(Y_{2};U)+H(Y_{1}|Y_{2},U).
\end{eqnarray}
Finally applying Fourier-Motzkin elimination to remove $R_{1}$, $R_{2}$ ($R=R_{1}+R_{2}$), $R^{'}$ and $R^{*}$ from
(\ref{app14.x1}), (\ref{zhenni4}), (\ref{zhenni5}), (\ref{app2}), (\ref{app4}), (\ref{app11}) and (\ref{app14}),
Theorem \ref{T2} is proved.

\textbf{Remarks}:  \textcolor[rgb]{1.00,0.00,0.00}{From the above proof, we see that the number of $U^{n}$ (see (\ref{app10}))
affects the secrecy rate. Then it is easy to explain why our new feedback scheme gains an advantage over Ahlswede and Cai's secret key based scheme \cite{AC},
and this is because our new scheme generates more $U^{n}$ (see (\ref{app4}), the number of $U^{n}$ is about $2^{nNI(Y_{1},V;U)}$)
than Ahlswede and Cai's scheme does
(the number of $U^{n}$ in Ahlswede and Cai's scheme is about $2^{nNI(Y_{1};U)}$).}

\section{The Non-degraded WTC-NF}\label{secIII}
\setcounter{equation}{0}

\begin{figure}[htb]
\centering
\includegraphics[scale=0.6]{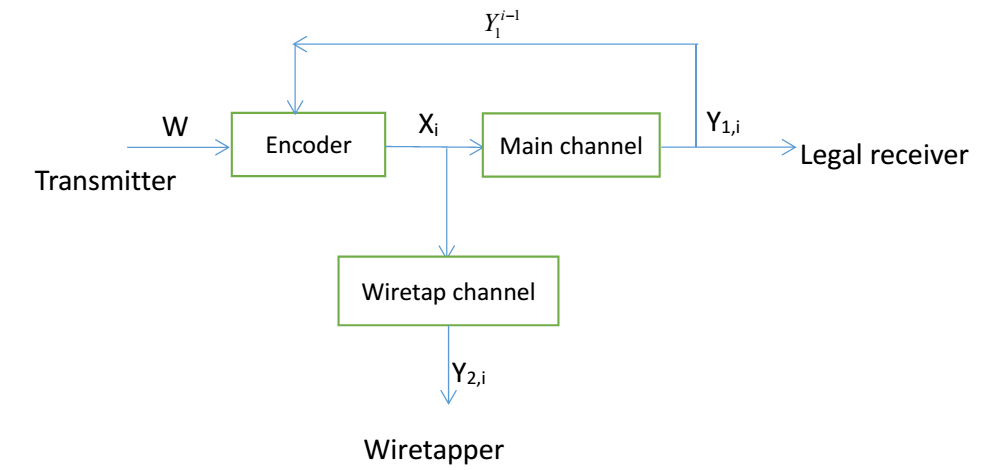}
\caption{The non-degraded WTC-NF}
\label{fx}
\end{figure}

The non-degraded WTC is the original basis for what now forms the compound WTC, which has been extensively investigated
in \cite{liang-1}-\cite{liang-x}. In order to study whether our new feedback scheme helps to increase the secrecy capacity of the
non-degraded WTC, and whether our new scheme takes advantage over the previous feedback scheme \cite{AC},
in this section, we study the non-degraded WTC-NF, see Figure \ref{fx}.

From Figure \ref{fx}, we know that for the non-degraded WTC-NF, given the channel input $X$,
the channel output $Y_{1}$ for the legal receiver is independent of the channel output $Y_{2}$
for the wiretapper, i.e., the Markov condition $Y_{1}\rightarrow X\rightarrow Y_{2}$ holds for the non-degraded WTC-NF.
Hence we see that the non-degraded WTC-NF is a special case of the WTC-NF described in Figure \ref{f1}.
For the non-degraded WTC-NF, its secrecy capacity is denoted as $C^{f*}_{s}$, and
lower and upper bounds on $C^{f*}_{s}$ are given as below.

\begin{theorem}\label{T2.x}
$C^{f*}_{s}\geq R^{**}_{s}$,
where
\begin{eqnarray}\label{trans1-gg}
&&R^{**}_{s}=\max_{P(v|x,y_{1})P(x)}\min\{[I(X;Y_{1},V)-I(X;Y_{2})]^{+}+H(Y_{1}|X),I(X;Y_{1})\},
\end{eqnarray}
the joint distribution is denoted by
\begin{eqnarray}\label{trans2-gg}
&&P(v,x,y_{1},y_{2})=P(v|x,y_{1})P(y_{1}|x)P(y_{2}|x)P(x),
\end{eqnarray}
and $V$ is bounded in cardinality by $|\mathcal{V}|\leq |\mathcal{X}|+2$.
\end{theorem}
\begin{IEEEproof}
Replacing $U$ by $X$, and along the lines of the proof of Theorem \ref{T2}, the rate
$$\max_{P(v|x,y_{1})P(x)}\min\{[I(X;Y_{1},V)-I(X;Y_{2})]^{+}+H(Y_{1}|X,Y_{2}),I(X;Y_{1})\}$$ is shown to be achievable. Then applying
the Markov condition $Y_{1}\rightarrow X\rightarrow Y_{2}$ to eliminate $Y_{2}$ in $H(Y_{1}|X,Y_{2})$, the secrecy rate $R^{**}_{s}$ is obtained.
\end{IEEEproof}

\begin{theorem}\label{T2.xx}
$C^{f*}_{s}\leq C^{f*-out}_{s}$,
where
\begin{eqnarray}\label{trans1-yyy}
&&C^{f*-out}_{s}=\max_{P(x)}\min\{H(Y_{1}|Y_{2}),I(X;Y_{1})\},
\end{eqnarray}
and the joint distribution is denoted by
\begin{eqnarray}\label{trans2-yy}
&&P(x,y_{1},y_{2})=P(y_{1}|x)P(y_{2}|x)P(x).
\end{eqnarray}
\end{theorem}
\begin{IEEEproof}
See Appendix \ref{rotk2}.
\end{IEEEproof}

Finally, replacing $U$ by $X$, along the proof of (\ref{e1.2}) (see \cite{AC}), and using
the Markov condition $Y_{1}\rightarrow X\rightarrow Y_{2}$ to eliminate $Y_{2}$ in $H(Y_{1}|X,Y_{2})$, an achievable secrecy rate $R^{non}_{s}$
of the non-degraded WTC-NF, which is constructed by Ahlswede and Cai's feedback coding scheme \cite{AC}, is given by
\begin{eqnarray}\label{e1.2.rri}
&&R^{non}_{s}=\max_{P(x)}\min\{[I(X;Y_{1})-I(X;Y_{2})]^{+}+H(Y_{1}|X),I(X;Y_{1})\}.
\end{eqnarray}
The comparison of $R^{non}_{s}$ and our new achievable secrecy rate $R^{**}_{s}$ will be given in the next section.

\section{Binary Example}\label{secIV}
\setcounter{equation}{0}

We consider a binary example. In this example, the channel consists of binary input and outputs satisfying
\begin{eqnarray}\label{e301}
&&Y_{1}=X\oplus Z_{1},\,\,\,Y_{2}=X\oplus Z_{2},
\end{eqnarray}
where $\oplus$ is the modulo addition over $\{0,1\}$, the noise $Z_{1}$ is $Bern(p_{1})$ ($p_{1}<0.5$), and the noise $Z_{2}$ is $Bern(p_{2})$ ($p_{2}<0.5$).
The noises $Z_{1}$ and $Z_{2}$ are independent of $X$, and they are mutually independent. From (\ref{e301}) and the properties of $Z_{1}$ and $Z_{2}$,
we can conclude that the Markov condition $Y_{1}\rightarrow X\rightarrow Y_{2}$ holds for this binary example, and hence
this binary model is a kind of the non-degraded WTC.

First, from \cite[Example 22.1]{network}, the secrecy capacity $C^{b}_{s}$ of the binary non-degraded WTC is
characterized by
\begin{eqnarray}\label{tvt1}
&&C_{s}^{b}=[h(p_{2})-h(p_{1})]^{+},
\end{eqnarray}
where $h(u)=-u\log(u)-(1-u)\log(1-u)$ and $[u]^{+}=\max\{0,u\}$.

Second, define $P(x=0)=\alpha$ and $P(x=1)=1-\alpha$ ($0\leq \alpha\leq 1$), then from (\ref{e301}) we have
\begin{eqnarray}\label{tvt1.eubla1}
&&P(Y_{1}=0)=\alpha\star p_{1}, \,\, P(Y_{1}=1)=\alpha\star (1-p_{1}),
\end{eqnarray}
\begin{eqnarray}\label{tvt1.eubla2}
&&P(Y_{2}=0)=\alpha\star p_{2}, \,\, P(Y_{2}=1)=\alpha\star (1-p_{2}),
\end{eqnarray}
where $u\star v=u(1-v)+(1-u)v$. Now substituting (\ref{tvt1.eubla1}), (\ref{tvt1.eubla2}) and (\ref{e301}) into
(\ref{e1.2.rri}), an achievable rate $C^{b-in}_{sf}$ of the binary non-degraded WTC-NF, which is constructed according to
\cite{AC}, is given by
\begin{eqnarray}\label{tvt1.eubla3}
&&C^{b-in}_{sf}=\max_{P(x)}\min\{[I(X;Y_{1})-I(X;Y_{2})]^{+}+H(Y_{1}|X),I(X;Y_{1})\}\nonumber\\
&&=\max_{\alpha}\min\{[h(\alpha\star p_{1})-h(p_{1})-h(\alpha\star p_{2})+h(p_{2})]^{+}+h(p_{1}),h(\alpha\star p_{1})-h(p_{1})\}\nonumber\\
&&\stackrel{(1)}=\min\{[h(p_{2})-h(p_{1})]^{+}+h(p_{1}),1-h(p_{1})\},
\end{eqnarray}
where (1) is achieved when $\alpha=\frac{1}{2}$.

Third, to calculate our new achievable rate $C^{b-in-new}_{sf}$ for the binary non-degraded WTC-NF,
define $P(x=0)=\alpha$ and $P(x=1)=1-\alpha$ ($0\leq \alpha\leq 1$), and further define
$P(v=0|x=0,y_{1}=0)=\gamma_{1}$, $P(v=1|x=0,y_{1}=0)=1-\gamma_{1}$, $P(v=0|x=0,y_{1}=1)=\gamma_{2}$, $P(v=1|x=0,y_{1}=1)=1-\gamma_{2}$,
$P(v=0|x=1,y_{1}=0)=\gamma_{3}$, $P(v=1|x=1,y_{1}=0)=1-\gamma_{3}$, $P(v=0|x=1,y_{1}=1)=\gamma_{4}$, $P(v=1|x=1,y_{1}=1)=1-\gamma_{4}$.
Then substituting (\ref{e301}) into Theorem \ref{T2.x}, our new
rate $C^{b-in-new}_{sf}$ is calculated by
\begin{eqnarray}\label{tvt7}
&&C^{b-in-new}_{sf}=\max_{\alpha,\gamma_{1},\gamma_{2},\gamma_{3},\gamma_{4}}\min\{[A-h(\alpha\star p_{2})+h(p_{2})]^{+}+h(p_{1}),h(\alpha\star p_{1})-h(p_{1})\}
\end{eqnarray}
where
\begin{eqnarray*}\label{tvt8}
&&A=\gamma_{1}\alpha(1-p_{1})\log\frac{\gamma_{1}(1-p_{1})}{\gamma_{1}\alpha(1-p_{1})+\gamma_{3}p_{1}(1-\alpha)}\nonumber\\
&&+(1-\gamma_{1})\alpha(1-p_{1})\log\frac{(1-\gamma_{1})(1-p_{1})}{(1-\gamma_{1})\alpha(1-p_{1})+(1-\gamma_{3})p_{1}(1-\alpha)}\nonumber\\
&&+\gamma_{2}p_{1}\alpha\log\frac{\gamma_{2}p_{1}}{\gamma_{2}p_{1}\alpha+\gamma_{4}(1-p_{1})(1-\alpha)}\nonumber\\
&&+(1-\gamma_{2})p_{1}\alpha\log\frac{(1-\gamma_{2})p_{1}}{(1-\gamma_{2})p_{1}\alpha+(1-\gamma_{4})(1-p_{1})(1-\alpha)}\nonumber\\
&&+\gamma_{3}p_{1}(1-\alpha)\log\frac{\gamma_{3}p_{1}}{\gamma_{3}p_{1}(1-\alpha)+\gamma_{1}(1-p_{1})\alpha}\nonumber\\
&&+(1-\gamma_{3})p_{1}(1-\alpha)\log\frac{(1-\gamma_{3})p_{1}}{(1-\gamma_{3})p_{1}(1-\alpha)+(1-\gamma_{1})(1-p_{1})\alpha}\nonumber\\
&&+\gamma_{4}(1-p_{1})(1-\alpha)\log\frac{\gamma_{4}(1-p_{1})}{\gamma_{4}(1-p_{1})(1-\alpha)+\gamma_{2}p_{1}\alpha}\nonumber\\
&&+(1-\gamma_{4})(1-p_{1})(1-\alpha)\log\frac{(1-\gamma_{4})(1-p_{1})}{(1-\gamma_{4})(1-p_{1})(1-\alpha)+(1-\gamma_{2})p_{1}\alpha}.
\end{eqnarray*}

Finally, to show the gap between the lower and upper bounds, we calculate the upper bound $C^{b-out}_{sf}$ on the secrecy capacity
of the binary non-degraded WTC-NF. Let $P(x=0)=\alpha$ and $P(x=1)=1-\alpha$ ($0\leq \alpha\leq 1$),
then from (\ref{e301}), we have
\begin{eqnarray}\label{tvt1.eubla1.rry}
&&P(Y_{1}=0,Y_{2}=0)=\alpha(1-p_{1})(1-p_{2})+(1-\alpha)pq,\nonumber\\
&&P(Y_{1}=0,Y_{2}=1)=\alpha(1-p_{1})p_{2}+(1-\alpha)p_{1}(1-p_{2}),\nonumber\\
&&P(Y_{1}=1,Y_{2}=0)=\alpha p_{1}(1-p_{2})+(1-\alpha)(1-p_{1})p_{2},\nonumber\\
&&P(Y_{1}=1,Y_{2}=1)=\alpha pq+(1-\alpha)(1-p_{1})(1-p_{2}).
\end{eqnarray}
Substituting (\ref{tvt1.eubla1.rry}), (\ref{tvt1.eubla1}), (\ref{tvt1.eubla2}) and (\ref{e301}) into
Theorem \ref{T2.xx}, the upper bound $C^{b-out}_{sf}$ for this binary case is given by
\begin{eqnarray}\label{tvt7.ggb}
&&C^{b-out}_{sf}=\max_{\alpha}\min\{B,h(\alpha\star p_{1})-h(p_{1})\},
\end{eqnarray}
where
\begin{eqnarray*}\label{tvt8.ggb}
&&B=(\alpha(1-p_{1})(1-p_{2})+(1-\alpha)pq)\log\frac{\alpha(1-p_{2})+(1-\alpha)p_{2}}{\alpha(1-p_{1})(1-p_{2})+(1-\alpha)pq}\nonumber\\
&&+(\alpha(1-p_{1})p_{2}+(1-\alpha)p_{1}(1-p_{2}))\log\frac{\alpha p_{2}+(1-\alpha)(1-p_{2})}{\alpha(1-p_{1})p_{2}+(1-\alpha)p_{1}(1-p_{2})}\nonumber\\
&&+(\alpha p_{1}(1-p_{2})+(1-\alpha)(1-p_{1})p_{2})\log\frac{\alpha(1-p_{2})+(1-\alpha)p_{2}}{\alpha p_{1}(1-p_{2})+(1-\alpha)(1-p_{1})p_{2}}\nonumber\\
&&+(\alpha pq+(1-\alpha)(1-p_{1})(1-p_{2}))\log\frac{\alpha p_{2}+(1-\alpha)(1-p_{2})}{\alpha pq+(1-\alpha)(1-p_{1})(1-p_{2})},
\end{eqnarray*}
and $\alpha\star p_{1}=\alpha(1-p_{1})+(1-\alpha)p_{1}$.

Numerical results for our new achievable rate $C^{b-in-new}_{sf}$,
Ahlswede and Cai's rate $C^{b-in}_{sf}$, the upper bound $C^{b-out}_{sf}$ on the secrecy capacity of the binary non-degraded WTC-NF,
and the secrecy capacity $C^{b}_{s}$ of the binary non-degraded WTC are given in the following Figs. \ref{f2}-\ref{f2-gg4}.
From these figures, we see that
our new scheme dominates the previous scheme \cite{AC} when $p_{2}$ is sufficiently small,
and the achievable rates of both schemes equal to the legal receiver's channel capacity $1-h(p_{1})$ when $p_{2}$ is increasing,
while this is impossible for the WTC without
feedback. \textcolor[rgb]{1.00,0.00,0.00}{Furthermore, as we see in these figures, the gap between our new achievable rate and Ahlswede and Cai's rate \cite{AC} is decreasing while
$p_{1}$ is increasing, which indicates that our new scheme gains more advantage over the previous scheme \cite{AC}
when the channel noise of the legitimate parties is small. Moreover, these figures show that the gap between the
upper and lower bounds is decreasing while $p_{1}$ is increasing. In Figure \ref{f2-gg3}, it is shown that for $p_{1}=0.1$,
our new scheme achieves the secrecy capacity of the binary non-degraded WTC-NF
($C^{b-in-new}_{sf}=C^{b-out}_{sf}$ for all $p_{2}$), and it performs better than the previous scheme \cite{AC}.
In Figure \ref{f2-gg4}, it is shown that when
$p_{1}$ goes up to $0.2$, both our new scheme and
the previous scheme \cite{AC} achieve the secrecy capacity of the binary non-degraded WTC-NF
($C^{b-in-new}_{sf}=C^{b-in}_{sf}=C^{b-out}_{sf}$ for all $p_{2}$), which equals to the main channel capacity $1-h(p_{1})$.}

\begin{figure}[htb]
\centering
\includegraphics[scale=0.5]{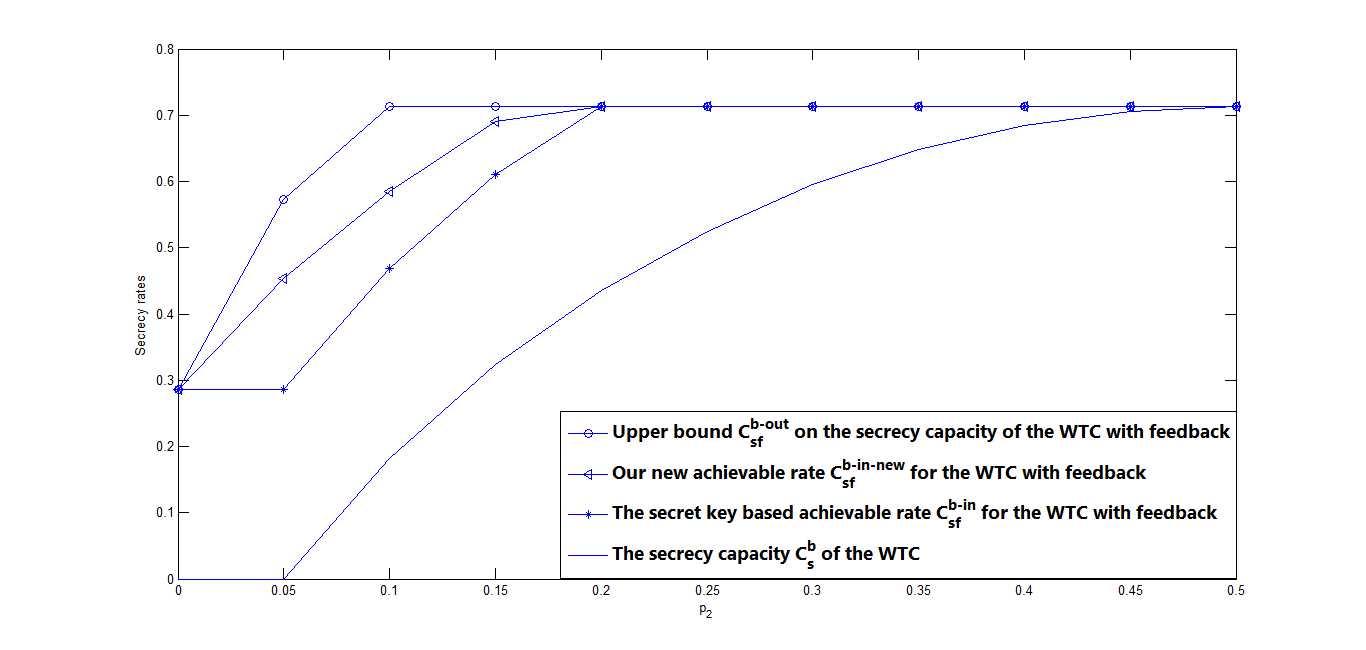}
\caption{Comparison of the bounds in (\ref{tvt1}), (\ref{tvt1.eubla3}), (\ref{tvt7}) and (\ref{tvt7.ggb})
for $p_{1}=0.05$ and several values of $p_{2}$}
\label{f2}
\end{figure}

\begin{figure}[htb]
\centering
\includegraphics[scale=0.5]{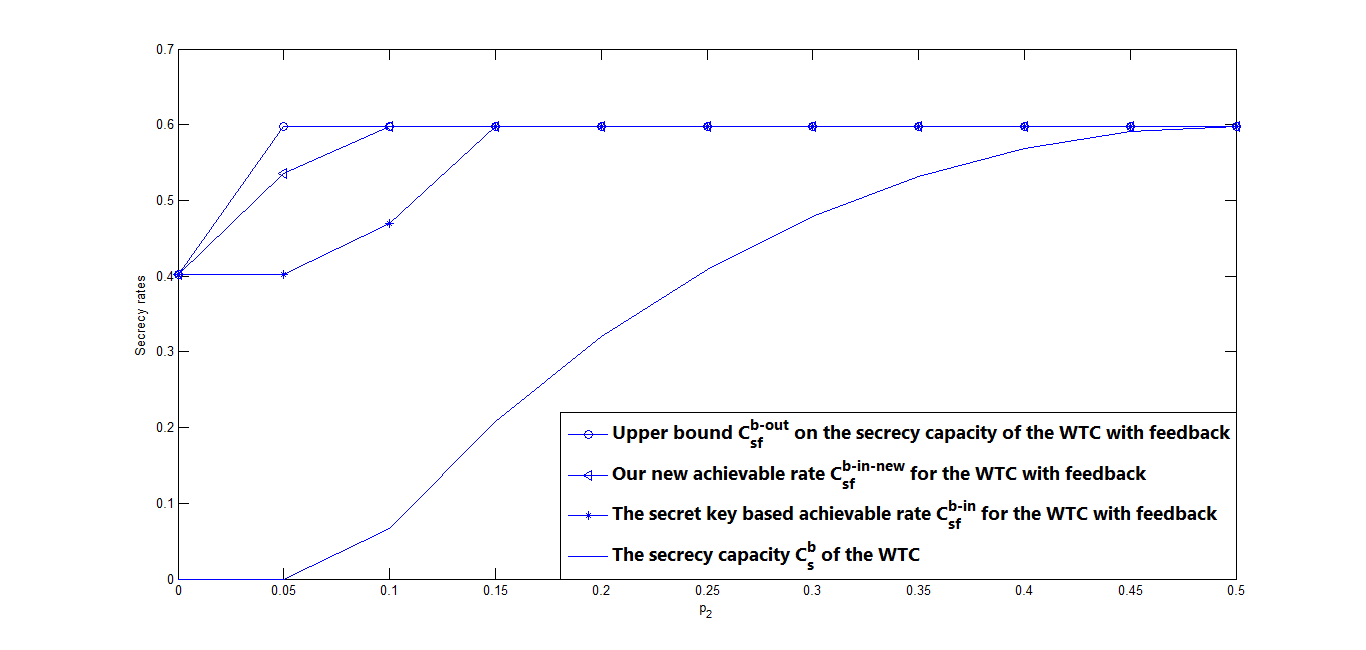}
\caption{\textcolor[rgb]{1.00,0.00,0.00}{Comparison of the bounds in (\ref{tvt1}), (\ref{tvt1.eubla3}), (\ref{tvt7}) and (\ref{tvt7.ggb})
for $p_{1}=0.08$ and several values of $p_{2}$}}
\label{f2-gg2}
\end{figure}

\begin{figure}[htb]
\centering
\includegraphics[scale=0.5]{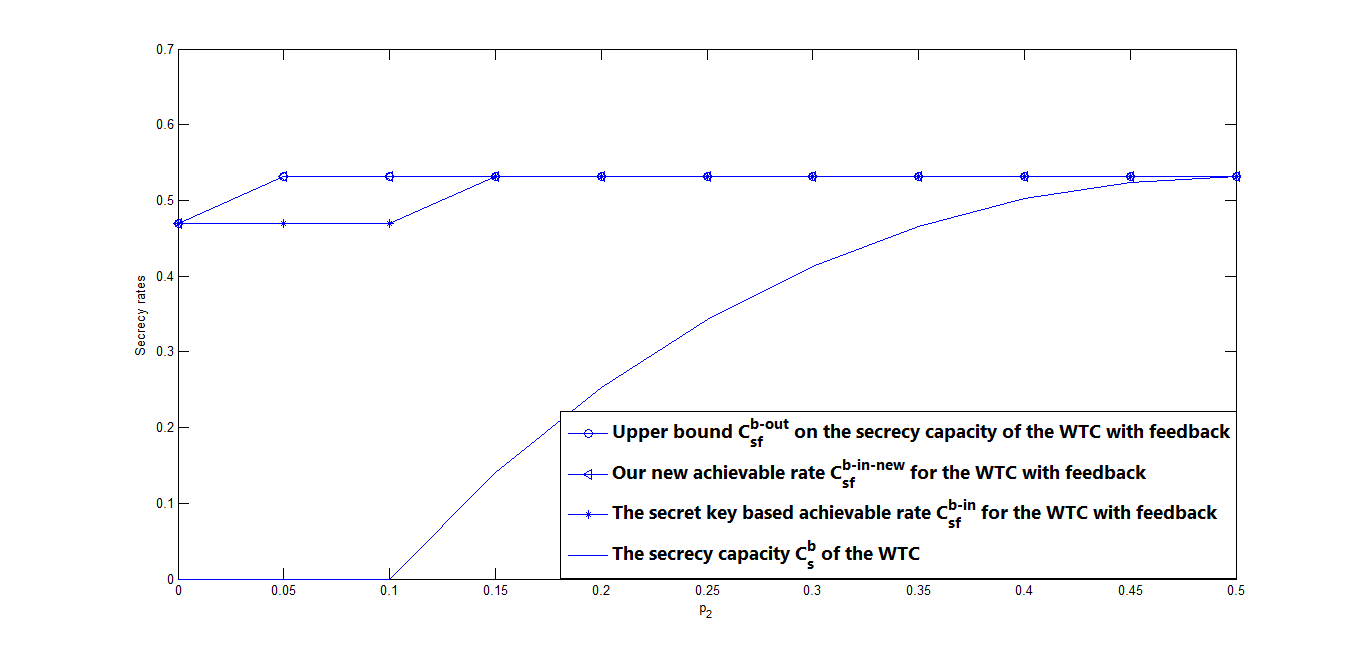}
\caption{\textcolor[rgb]{1.00,0.00,0.00}{Comparison of the bounds in (\ref{tvt1}), (\ref{tvt1.eubla3}), (\ref{tvt7}) and (\ref{tvt7.ggb})
for $p_{1}=0.1$ and several values of $p_{2}$}}
\label{f2-gg3}
\end{figure}

\begin{figure}[htb]
\centering
\includegraphics[scale=0.5]{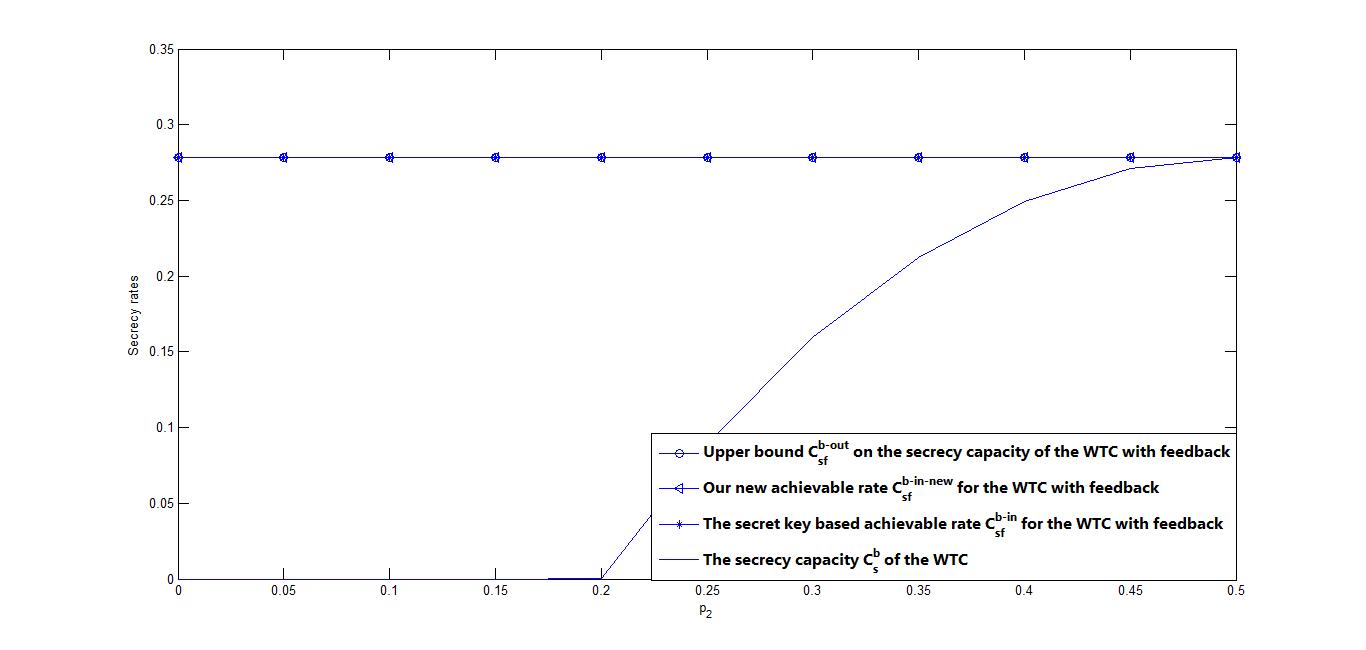}
\caption{\textcolor[rgb]{1.00,0.00,0.00}{Comparison of the bounds in (\ref{tvt1}), (\ref{tvt1.eubla3}), (\ref{tvt7}) and (\ref{tvt7.ggb})
for $p_{1}=0.2$ and several values of $p_{2}$}}
\label{f2-gg4}
\end{figure}

\section{Conclusion\label{secV}}

A new feedback strategy of the WTC-NF is proposed, and it is
shown to be better than the already existing feedback strategy for the WTC-NF.
The result of this paper is further illustrated by a binary symmetric case, and it is shown that
the advantage of this new feedback strategy is highlighted
when the channel noises are small.

\section*{Acknowledgment}

The authors are very grateful to Prof. Yingbin Liang and the anonymous reviewers for the helpful comments on improving this paper.

\renewcommand{\theequation}{A\arabic{equation}}
\appendices\section{Proof of Theorem \ref{T2.xx}\label{rotk2}}
\setcounter{equation}{0}

First, define
\begin{eqnarray}\label{b1}
&&X\triangleq X_{J},\,\,Y_{1}\triangleq Y_{1,J},\,\, Y_{2}\triangleq Y_{2,J},
\end{eqnarray}
where $J$ is a time sharing RV
uniformly distributed in $\{1,2,...,N\}$ and it is independent of the random vectors $X^{N}$, $Y_{1}^{N}$ and $Y_{2}^{N}$.

Then, notice that
\begin{eqnarray}\label{b2}
&&R-\epsilon\stackrel{(1)}\leq\frac{1}{N}H(W)\nonumber\\
&&=\frac{1}{N}(I(W;Y_{1}^{N})+H(W|Y_{1}^{N}))\nonumber\\
&&\stackrel{(2)}\leq \frac{1}{N}(I(W;Y_{1}^{N})+\delta(P_{e}))\nonumber\\
&&=\frac{1}{N}\sum_{i=1}^{N}(H(Y_{1,i}|Y_{1}^{i-1})-H(Y_{1,i}|W,Y_{1}^{i-1}))+\frac{\delta(P_{e})}{N}\nonumber\\
&&\leq \frac{1}{N}\sum_{i=1}^{N}(H(Y_{1,i})-H(Y_{1,i}|W,Y_{1}^{i-1},X_{i}))+\frac{\delta(P_{e})}{N}\nonumber\\
&&\stackrel{(3)}=\frac{1}{N}\sum_{i=1}^{N}(H(Y_{1,i})-H(Y_{1,i}|X_{i}))+\frac{\delta(P_{e})}{N}\nonumber\\
&&\stackrel{(4)}=\frac{1}{N}\sum_{i=1}^{N}(H(Y_{1,i}|J=i)-H(Y_{1,i}|X_{i},J=i))+\frac{\delta(P_{e})}{N}\nonumber\\
&&\stackrel{(5)}=H(Y_{1,J}|J)-H(Y_{1,J}|X_{J},J)+\frac{\delta(P_{e})}{N}\nonumber\\
&&\stackrel{(6)}\leq H(Y_{1,J})-H(Y_{1,J}|X_{J})+\frac{\delta(\epsilon)}{N}\nonumber\\
&&\stackrel{(7)}=I(X;Y_{1})+\frac{\delta(\epsilon)}{N},
\end{eqnarray}
where (1) is due to (\ref{e205-1}), (2) is implied by Fano's lemma,
(3) is due to the main channel is discrete memoryless, i.e.,
the Markov condition $(W,Y_{1}^{i-1})\rightarrow X_{i}\rightarrow Y_{1,i}$ holds,
(4) is due to $J$ is independent of the random vectors $X^{N}$, $Y_{1}^{N}$ and $Y_{2}^{N}$,
(5) is due to $Pr\{J=i\}=\frac{1}{N}$ for $1\leq i\leq N$,
(6) is implied by the Markov chain
$J\rightarrow X_{J}\rightarrow Y_{1,J}$, $P_{e}\leq \epsilon$ and $\delta(P_{e})$ is a monotonic increasing function of $P_{e}$,
and (7) is implied by (\ref{b1}).
Letting $\epsilon\rightarrow 0$ (here note that $\delta(\epsilon)\rightarrow 0$ as $\epsilon\rightarrow 0$), we get $R\leq I(X;Y_{1})$.

Moreover, also notice that
\begin{eqnarray}\label{b3}
&&R-\epsilon\stackrel{(a)}\leq\frac{1}{N}H(W|Y_{2}^{N})\nonumber\\
&&=\frac{1}{N}(H(W|Y_{2}^{N})+H(W|Y_{2}^{N},Y_{1}^{N})-H(W|Y_{2}^{N},Y_{1}^{N}))\nonumber\\
&&\stackrel{(b)}\leq\frac{1}{N}(H(W|Y_{2}^{N})+\delta(P_{e})-H(W|Y_{2}^{N},Y_{1}^{N}))\nonumber\\
&&=\frac{1}{N}(I(W;Y_{1}^{N}|Y_{2}^{N})+\delta(P_{e}))\nonumber\\
&&\leq\frac{1}{N}\sum_{i=1}^{N}H(Y_{1,i}|Y_{2,i})+\frac{\delta(P_{e})}{N}\nonumber\\
&&\stackrel{(c)}=\frac{1}{N}\sum_{i=1}^{N}H(Y_{1,i}|Y_{2,i},J=i)+\frac{\delta(P_{e})}{N}\nonumber\\
&&\stackrel{(d)}=H(Y_{1,J}|Y_{2,J},J)+\frac{\delta(P_{e})}{N}\nonumber\\
&&\stackrel{(e)}\leq H(Y_{1,J}|Y_{2,J})+\frac{\delta(\epsilon)}{N}\nonumber\\
&&\stackrel{(f)}=H(Y_{1}|Y_{2})+\frac{\delta(\epsilon)}{N},
\end{eqnarray}
where (a) is implied by (\ref{e205-2}),
(b) is implied by Fano's lemma,
(c) is implied by $J$ is independent of the random vectors $X^{N}$, $Y_{1}^{N}$ and $Y_{2}^{N}$, (d) is implied by
$Pr\{J=i\}=\frac{1}{N}$ for $1\leq i\leq N$,
(e) is implied by $P_{e}\leq \epsilon$ and $\delta(P_{e})$ is a monotonic increasing function of $P_{e}$,
and (f) is implied by (\ref{b1}).
Letting $\epsilon\rightarrow 0$, we get $R\leq H(Y_{1}|Y_{2})$.

Combining (\ref{b2}) and (\ref{b3}), we have
$R\leq \min\{I(X;Y_{1}), H(Y_{1}|Y_{2})\}$. Here note that both the characters $I(X;Y_{1})$ and $H(Y_{1}|Y_{2})$ are functions of $P(x)$, then
choosing a $P(x)$ to maximize $\min\{I(X;Y_{1}), H(Y_{1}|Y_{2})\}$, we get $R\leq \max\min\{I(X;Y_{1}), H(Y_{1}|Y_{2})\}$.
The proof of Theorem \ref{T2.xx} is completed.

\end{document}